\documentclass[useAMS,usenatbib]{mn2e}

\usepackage{epsfig}
\usepackage{graphicx}
\usepackage{amssymb}

\title{ISW measurements with photometric redshift surveys: 2MASS results
  and future prospects}
\author[C. L. Francis and J. A. Peacock]
       {{Caroline L. Francis and John A. Peacock\thanks{E-mail: jap@roe.ac.uk}} \\
SUPA\thanks{Scottish Universities Physics
	     Alliance} Institute for Astronomy, University of
  Edinburgh, Royal Observatory, Blackford Hill, Edinburgh EH9 3HJ, UK}

\date{}

\pagerange{\pageref{firstpage}--\pageref{lastpage}}
\pubyear{2009}

\def\[{\begin{equation}}
\def\]{\end{equation}}
\def\gsim{\mathrel{\lower0.6ex\hbox{$\buildrel {\textstyle >}
 \over {\scriptstyle \sim}$}}}
\def\lsim{\mathrel{\lower0.6ex\hbox{$\buildrel {\textstyle <}
 \over {\scriptstyle \sim}$}}}

\begin{document}

\maketitle

\label{firstpage}

\begin{abstract}
\noindent In a flat universe dominated by dark energy, the Integrated
Sachs-Wolfe (ISW) effect can be detected as a large-angle
cross-correlation between the CMB and a tracer of large scale
structure. We investigate whether the inconclusive ISW signal derived
from 2MASS galaxy maps can be improved upon by including photometric
redshifts for the 2MASS galaxies. These redshifts are derived by
matching the 2MASS data with optical catalogues generated from
SuperCOSMOS scans of major photographic sky surveys. We find no
significant ISW signal in this analysis; an ISW effect of the form
expected in a $\Lambda$CDM universe is only weakly preferred over no
correlation, with a likelihood ratio of 1.5:1. We consider ISW
detection prospects for future large scale structure surveys with
fainter magnitude limits and greater survey depth; even with the best
possible data, the ISW cross-correlation signal would be expected to
evade detection in $\gtrsim 10\%$ of cases.

\end{abstract}


\section{Introduction}

\label{sec:intro}

It is now widely accepted that the universe today is dominated by Dark
Energy. Evidence for this comes from a number of sources: supernovae
observations indicate a luminosity distance-redshift relation
consistent with accelerated expansion
\citep{Reiss_MNRAS,Perlmutter_MNRAS}; the combination of CMB and
galaxy clustering measurements places tight constraints on the Dark
Energy density \citep{GPE_LAM90,GPE_LAM02,Spergel_WMAP3} and Baryon
Acoustic Oscillation (BAO) measurements suggest an angular diameter
distance-redshift relation consistent with Dark Energy
\citep{Eisenstein_BAO_MNRAS,WJP_BAO07}.

Further independent evidence for Dark Energy can be provided by
detection of the Integrated Sachs-Wolfe (ISW) effect, a gravitational
secondary anisotropy in the CMB. The ISW effect is a net change in the
energy of a CMB photon as it passes through evolving gravitational
potential wells. The ISW temperature fluctuations introduced via the
ISW effect can be calculated from
\[
\frac{\Delta T^{\textrm{\tiny ISW}}}{T_{\textrm{\tiny CMB}}} =
2\int_{t_{\textrm{\tiny
LS}}}^{t_{0}}{\frac{\dot{\Phi}(\vec{x}(t),t)}{c^{2}}} \, \mathrm{d}t
\label{eq:isw},
\]
where $t_{0}$ and $t_{\textrm{\tiny LS}}$ denote the times
today and at last scattering respectively; $\vec{x}$ is the position
along the line of sight of the photon at time $t$ and $\Phi$ is the
gravitational potential \citep{Martinez-Gonzalez_Sanz_Silk}.  

The ISW effect is a large scale phenomenon: temperature fluctuations
tend to cancel on smaller scales as photons pass through many
potential wells \citep{Hu_Dodelson_CMB}. The ISW effect is
sub-dominant to the primary CMB anisotropies and exists on scales
where cosmic variance is large, making its detection from CMB data
alone infeasible. Instead, the cross-correlation between CMB maps and
tracers of large scale structure is used
\citep{Crittenden_Turok_1996}. Many different tracers have been
correlated e.g. X-ray \citep{Boughn_Crittenden_2005}; Sloan Digital
Sky Survey \citep{Fosalba,Scranton_2003}; 
Luminous Red Galaxies \citep{Granett_ISW09,Sawangwit_ISW10};
radio \citep{Nolta,Pietrobon_ISW06}; and quasars \citep{Giannantonio_2006}. 
Not all of these lead to significant signals, but the
strongest claimed detections currently lie at the $3\sigma-4\sigma$
level. \citet{Giannantonio_2008} cross-correlated data from
a number of different large scale structure surveys with WMAP CMB data
and reported an overall ISW effect that was significant at the
$4.5\sigma$ level.

\citet{Afshordi} cross-correlated the Two Micron All Sky Survey
(2MASS, \citealp{Jarrett_2MASS}) and year-1 WMAP data
\citep{Bennett_WMAP1_MNRAS} and claimed to detect the ISW effect at
$2.5\sigma$. However, no significant effect was found in the updated
work of \citet{Rassat07} (hereafter R07), who correlated 2MASS with the
year-3 WMAP CMB data \citep{Hinshaw_WMAP3}. In the present
paper, we investigate whether the addition of photometric redshifts to
the 2MASS survey data can boost the significance of any ISW signal. We
also consider detection prospects for similar surveys with fainter
magnitude limits and surveys with greater depth.

For ease of comparison with R07 we adopt the same cosmological model,
namely a flat universe with $\Omega_{m} = 0.3$, $h = 0.7$ where the
Hubble parameter is \mbox{$H_{0}=$100 $h$ kms$^{-1}$Mpc$^{-1}$},
spectral index $n=1$, $\Omega_{b} = 0.05$ and $\sigma_{8} =
0.75$. Section \ref{sec:data} describes the data that we use in this
paper and details of the masks for the galaxy and CMB maps. Section
\ref{sec:Theory_bias} outlines the theoretical cross-correlation
signal expected in a $\Lambda$CDM universe and Section
\ref{sec:cross-corr} presents the results of our harmonic space
cross-correlation analysis. Section \ref{sec:stat_power} assesses the
statistical power of the method used here as a detection tool, and
investigates the detection prospects of more ambitious large scale
structure surveys. Our conclusions are presented in Section
\ref{sec:conclusions}.


\section{Data}

\label{sec:data}

The large scale structure dataset used in the cross-correlation
analysis is by far the greatest source of uncertainty. On the scales
of the ISW effect, the CMB data are signal dominated
\citep{Spergel_WMAP3} whilst the galaxy data are afflicted by shot
noise, uncertainties in the bias relation and photometric redshift
errors.

\subsection{2MASS galaxy data}

2MASS is an all-sky survey in the \textit{J}, \textit{H} and
\textit{K}$_{\textrm{s}}$ bands. The final extended source catalogue
(XSC) contains over 1.6 million objects, over 98\% of which are
galaxies \citep{Jarrett_2MASS}. This dataset is ideal for constructing
maps of galaxy mass density since near infrared selection means that
2MASS is sensitive to old stars and hence the most massive structures.

Photometric redshifts for the 2MASS XSC have been generated by
matching the 2MASS data with optical catalogues generated from
SuperCOSMOS scans of the major photographic sky surveys (UKST in the
south; POSS2 in the north). Details of the SuperCOSMOS catalogue
construction process are given in \citet{Hambly_superCOSMOS}. These
catalogues have been given an accurate photometric calibration using a
mixture of SDSS photometry, plate overlaps, and requiring uniformity
in average colour between optical band and 2MASS $J$; the details of
the process are described in \citet{Peacock_photoz}. The photographic
data define an all-sky optical galaxy catalogue in photographic
$B$,$R$, $I$ bands, reaching completeness limits of approximately
$B=21.5$, $R=20.5$, $I=19$. Since the $I$-band plates are
significantly less deep, they were not used in the current analysis,
leaving a 5-band $BRJHK$ dataset from which photometric redshifts were
to be derived. Owing largely to SDSS, 2dFGRS, and 6dFGS, a very
substantial set of calibrating spectroscopy exists -- for about 30\%
of the catalogue.  It was therefore possible to take a highly direct
approach to the generation of photometric redshifts, averaging over
neighbours of known redshift at a given location in the 5-band
magnitude space. By using magnitudes rather than colours, this
automatically builds in information from the luminosity function, so
that bright galaxies are never allocated an extreme redshift that
would require them to have an unrealistic luminosity. As a result, the
scatter in photometric redshift declines towards $z=0$, as shown in
Fig.~1. The overall rms in $z_{\rm phot}-z$ is 0.033. By construction,
this process yields photometric redshifts that are unbiased in that
the mean true redshift at given $z_{\rm phot}$ is equal to $z_{\rm phot}$.
An inevitable consequence is that there is then a bias in $z_{\rm phot}$
at given true $z$: $z_{\rm phot}$ is overestimated near $z=0$ and
underestimated at high $z$, by an offset of 0.02 to 0.03. In any case,
what we will need below is the ability to predict the distribution of
true $z$ from a band of $z_{\rm phot}$; provided this is known, any
bias is unimportant.

The optical
filter properties in UKST and POSS2 are very slightly different, so
the process was performed separately in the two hemispheres. However,
there is deeper calibrating spectroscopy in the north, which
introduces the possibility of a small bias. We therefore made a small
rescaling as a function of redshift so that the global redshift
distributions in each hemisphere were forced to be identical (in
shape, but not in number) for our standard extinction-corrected
$K$-selected galaxy sample.

\begin{figure}
\centering
\rotatebox{270}{
\epsfig{file=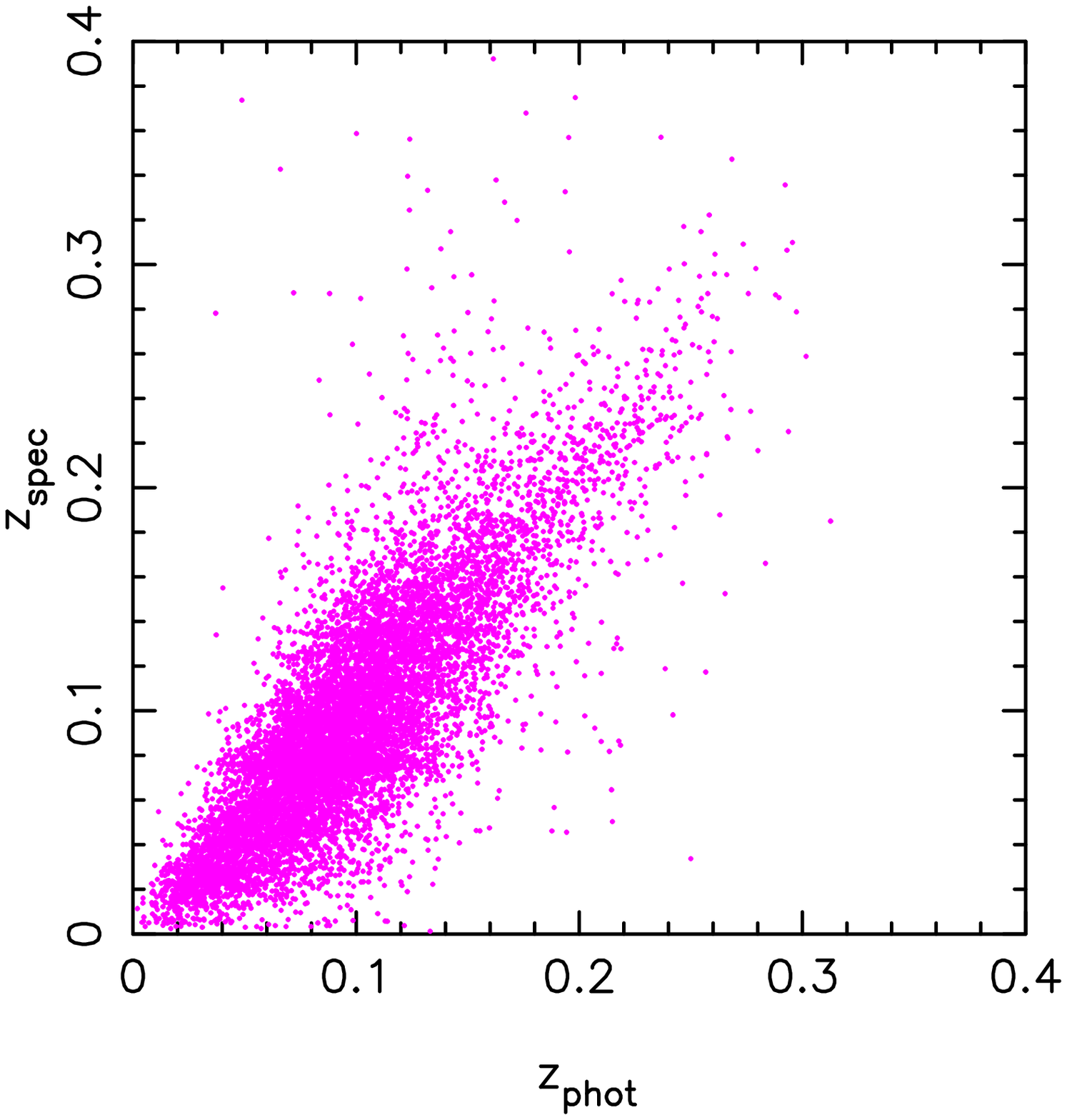,width=8.0cm,height=12.0cm}
}
\caption{A comparison of photometric and spectroscopic redshifts for
2MASS galaxies. The photometric redshifts are computed from the BRJHK
dataset resulting from combining 2MASS with SuperCOSMOS optical
photometry. \label{fig:zpzs}}
\end{figure}

\begin{figure}
\centering
\rotatebox{270}{
\epsfig{file=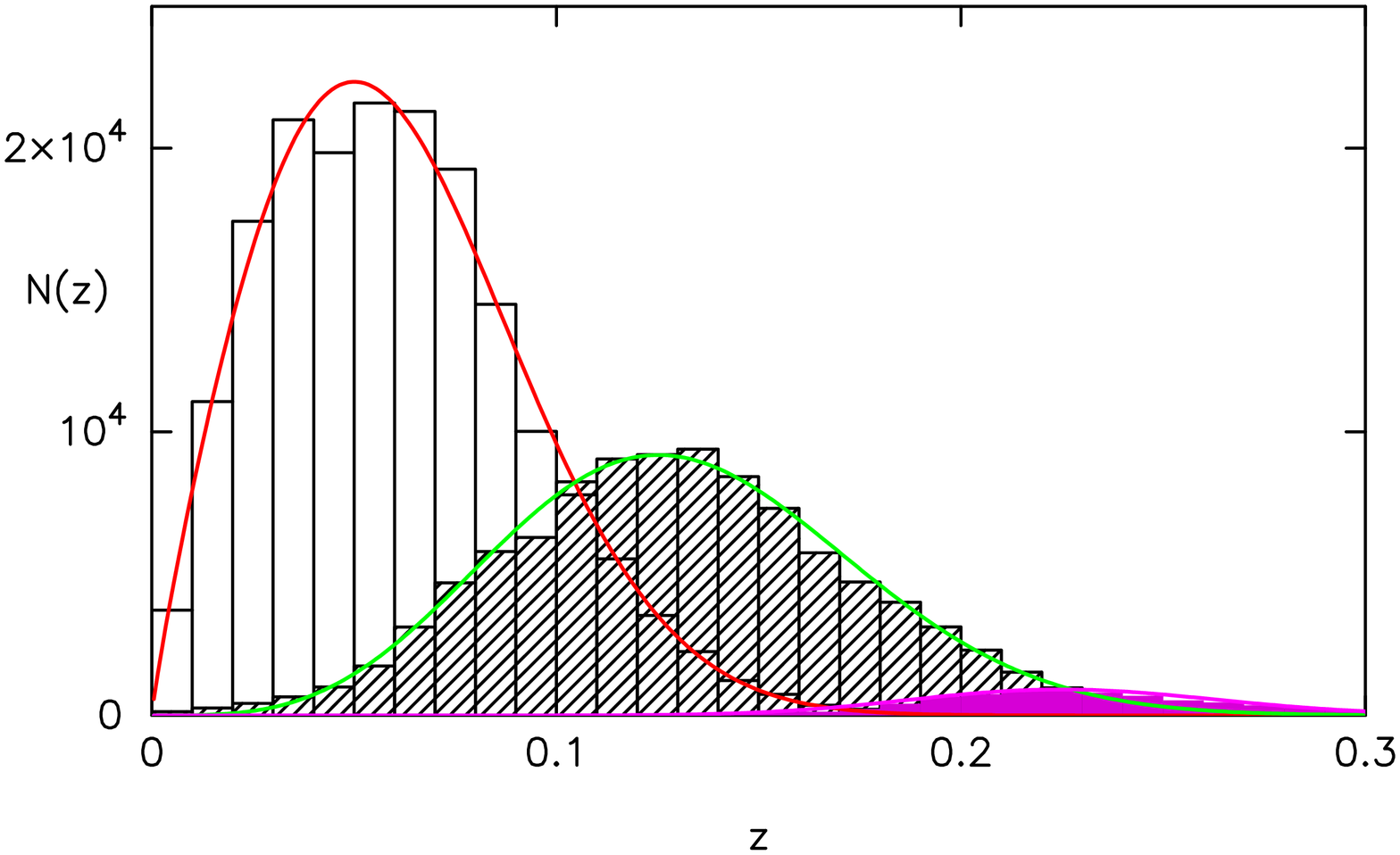,width=6.0cm,height=8.0cm}
}
\caption{The redshift distribution of galaxies selected by photometric
  redshift to lie in slices $z=0.0-0.1$, $z=0.1-0.2$ and
  $z=0.2-0.3$. The histograms show the distribution of galaxies in the
  photometric redshift shells with spectroscopic information; here the
  redshift distribution functions are normalized to match these
  data. \label{fig:zdistns}}
\end{figure}

We construct 3 galaxy subsamples in approximately independent
photometric redshift slices of thickness $\Delta z = 0.1$ out to
$z=0.3$. The true redshift distribution of galaxies in each of these
slices is calculated using the calibrating spectroscopic information;
these are shown in Fig. \ref{fig:zdistns}. We parametrize the redshift
distributions using
\[
\frac{\mathrm{d}N}{\mathrm{d}z} \propto z^{\alpha}\exp\left\{-(z/z_{*})^{\beta}\right\}, \label{eq:dNdzfit}
\]
and perform a least squares fit to obtain the distribution
parameters. The best-fitting values of $\alpha$, $\beta$ and $z_{*}$
for each photometric redshift slice are given in Table
\ref{table:galfacts}. It is clear that the $0.2 < z < 0.3$ slice makes
very little contribution to the 2MASS sky distribution; yet by virtue
of its larger volume, this slice will double the total ISW signal
compared to a limit at $z<0.2$. We may thus hope that the use of
redshift information in the current study will improve the chances of
detection.

\begin{table}
\centering
\begin{tabular}{cccccc} \hline
Slice & $N_{\textrm{\scriptsize gals}}$ & $N_z$ & $\alpha$ & $\beta$ & $z_{*}$ \\ \hline
$0.0<z<0.1$ & 451329 & 195789  & 0.88 & 2.24 & 0.076 \\
$0.1<z<0.2$ & 258673 & 101912  & 3.46 & 2.27 & 0.104 \\
$0.2<z<0.3$ & 19162 & 8543 & 17.4 & 2.12 & 0.083 \\ \hline
\end{tabular}
\caption{The redshift distribution of galaxies in each photometric
  redshift slice is described by equation (\ref{eq:dNdzfit}). The
  table gives the total number of galaxies in each slice,
  the number having spectroscopic redshifts, and best-fitting values of
  $\alpha$, $\beta$ and $z_{*}$ for each photometric redshift
  slice. \label{table:galfacts}}
\end{table}

It is necessary to mask the region of sky around the Galactic plane
before making the galaxy overdensity maps, since the survey is
incomplete here and confusion is more likely. Following R07, we use
the dust maps of \citet{Schlegel_dust} to mask regions with $K$-band
reddening $A_{K}>0.05$, which leaves approximately 67\% of the sky
unmasked. For the cross-correlation analysis we impose magnitude cuts
$12.0<K<13.8$ on the extinction corrected 2MASS data to ensure uniform
coverage across the whole sky. The number of galaxies remaining in
each of the photometric redshift slices is also given in Table
\ref{table:galfacts}.

\begin{figure*}

\centering
\epsfig{file=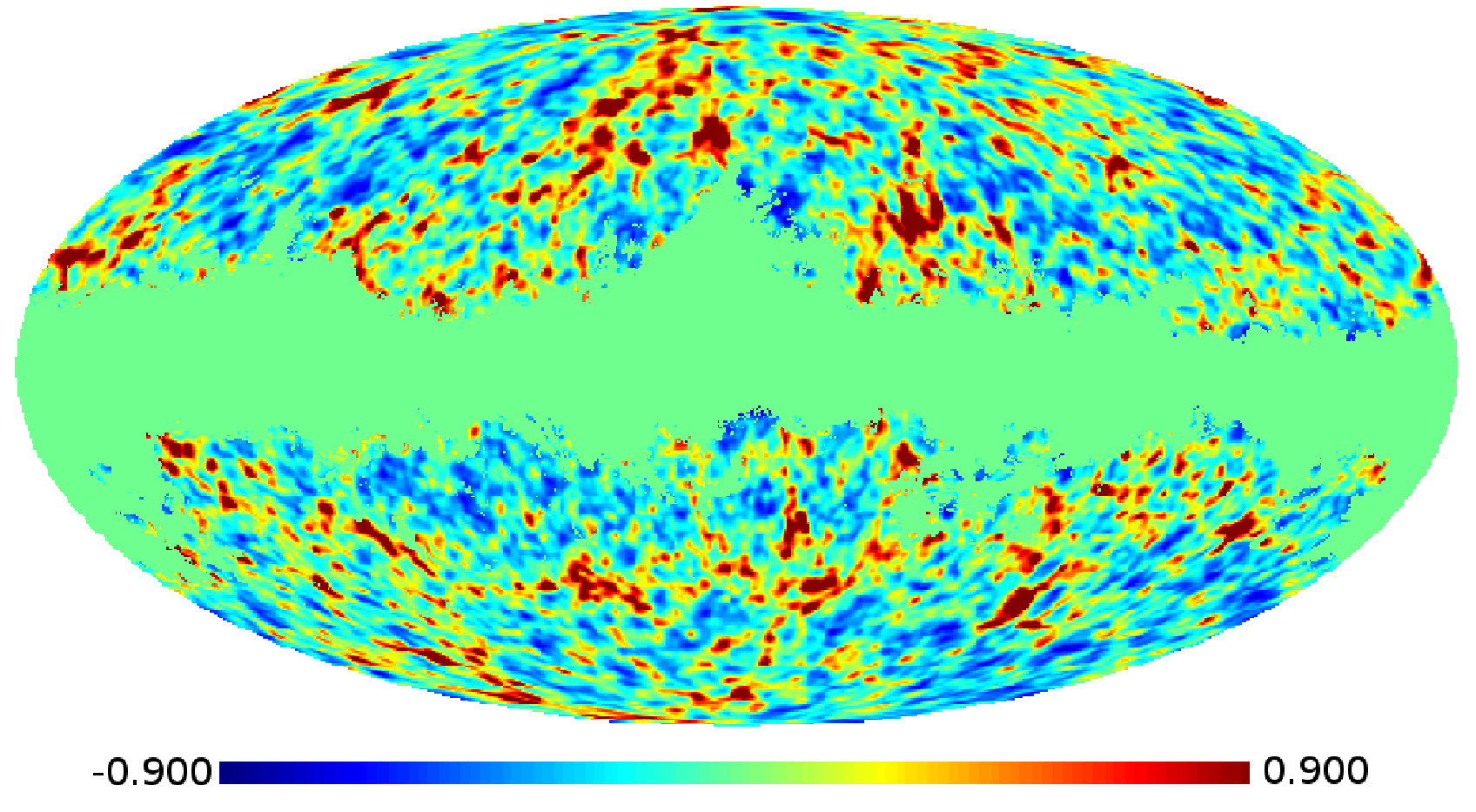,width=14.0cm}
\epsfig{file=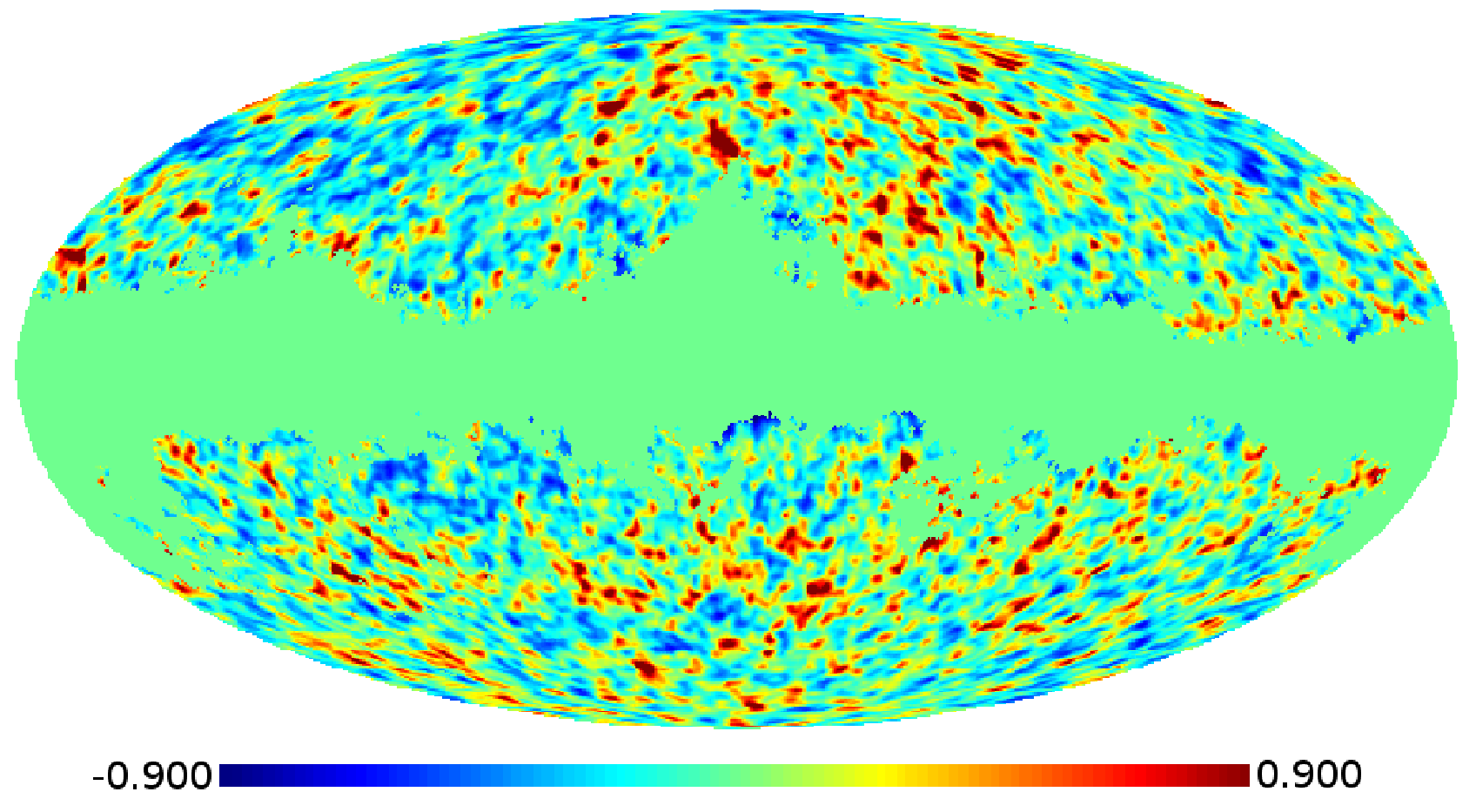,width=14.0cm}
\epsfig{file=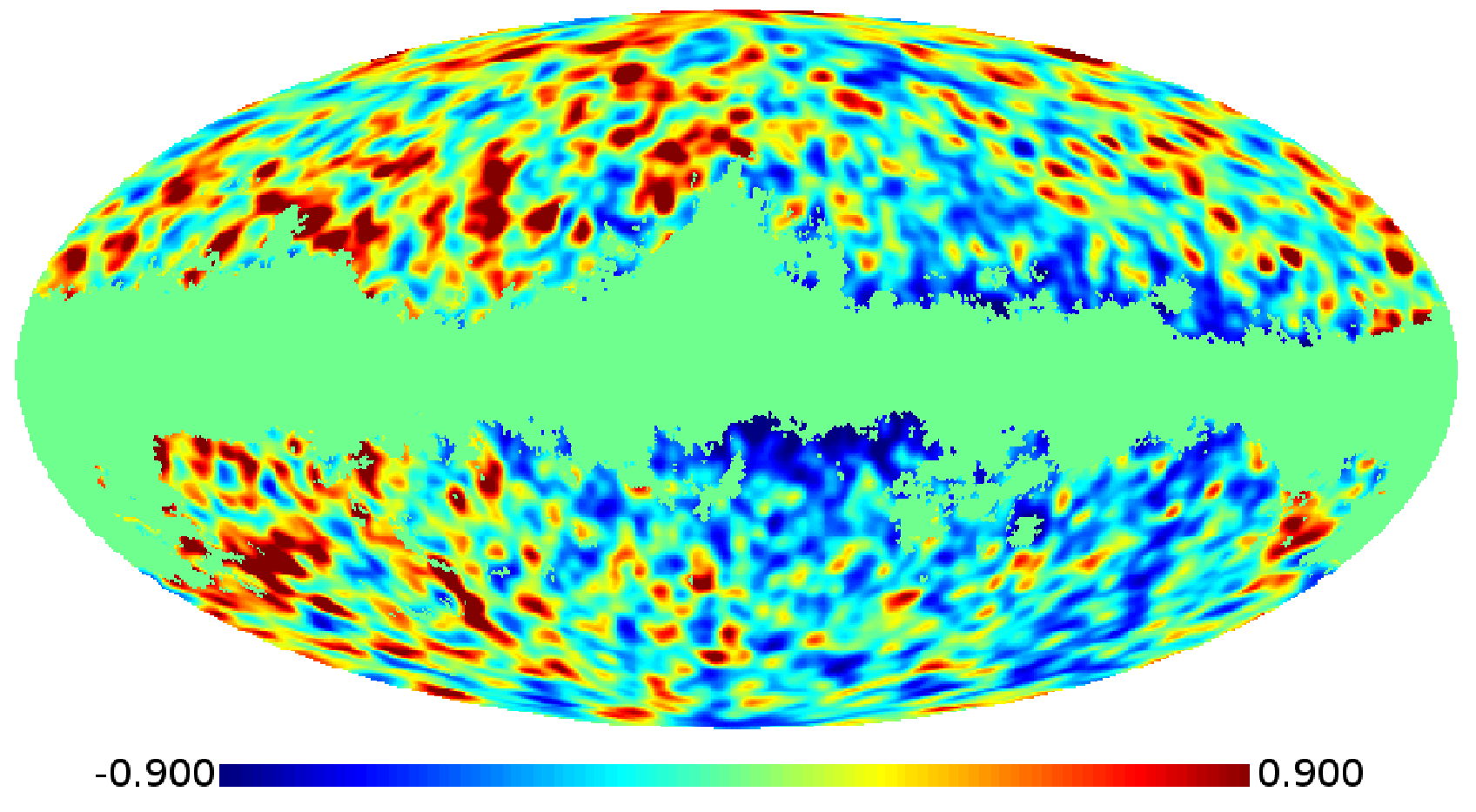,width=14.0cm}
\caption{Maps of the 2MASS galaxy overdensity in each of the 3
  redshift slices considered: $0.0<z<0.1$ (top), $0.1<z<0.2$ (middle)
  and $0.2<z<0.3$ (bottom). The density field is smoothed with a
  Gaussian of FWHM = $100'$ for the $z=0.0-0.1$ and $z=0.1-0.2$ slices
  and of FWHM = $200'$ for the $z=0.2-0.3$ slice. \label{fig:2mass_maps}}
\end{figure*}

We use HEALPix software (G\'orski et al. 2005; see also
\texttt{http://healpix.jpl.nasa.gov}) to generate maps of the galaxy
overdensity and to compute the spherical harmonic coefficients
$a_{\ell m}$ used in the cross-correlation
analysis. Fig. \ref{fig:2mass_maps} shows the galaxy overdensity in
each of our photometric redshift slices.

\subsection{WMAP CMB data}

We take our CMB temperature data from the third year WMAP results
\citep{Hinshaw_WMAP3}. Cross-correlations are computed for the 3
bands least contaminated by Galactic foregrounds ($Q$, $V$ and $W$)
and also for the Internal Linear Combination (ILC) map. To minimise
the effects of Galactic foregrounds we use the \emph{foreground
reduced} sky maps in which models for synchrotron, free-free and dust
emission have been subtracted. It is however prudent to mask the WMAP
maps at low Galactic latitudes and the WMAP Kp2 mask is used to this
end, leaving $\sim 85\%$ of the sky unmasked.

The fact that both the galaxy maps and the CMB maps are masked will
mean that spherical harmonic coefficients measured from the maps are
not exact and an estimator needs to be used. Here, following
R07,  we compensate for the lack of sky coverage using
the factor $f_{\textrm{\scriptsize sky}}$ in equation (\ref{eq:cgtcomp}).

We now summarise the necessary theory of the ISW effect (see
e.g. Cooray et al. 2002, Afshordi et al. 2004 for details).


\section{Theory and galaxy bias}

\label{sec:Theory_bias}

Expanding the projected galaxy density field in a given redshift shell
and the ISW temperature fluctuations in spherical harmonics, the cross
correlation between the galaxy overdensity field and the CMB
temperature field is:
\[
\left\langle a_{\ell m}^{\textrm{\scriptsize g}} a_{\ell'
m'}^{\textrm{\scriptsize T} *} \right\rangle = C_{\textrm{\scriptsize
gT}}(\ell)\delta_{\ell \ell'}^{\textrm{\scriptsize K}}\delta_{m
m'}^{\textrm{\scriptsize K}},
\]
where $a_{\ell m}^{\textrm{\scriptsize g}}$ and $a_{\ell
m}^{\textrm{\scriptsize T}}$ are the spherical harmonic coefficients
for the galaxy overdensity and temperature fields respectively and
$\delta^{\textrm{\scriptsize K}}$ is the Kronecker delta.

In Fourier space and within the linear regime, Poisson's equation
relates the time derivative of the potential field to the matter
density field:
\[
k^{2}\dot{\Phi}_{k} =
\frac{-3H_{0}^{2}\Omega_{m}}{2}\frac{\mathrm{d}}{\mathrm{d}t}\left[\frac{g(a)}{a}\right]\delta_{k}(z=0)
\label{eq:poisson},
\]
where $g(a)$ is the linear growth factor, normalised to $g=1$
today. From the definition of the ISW temperature fluctuations we then
have:
\begin{eqnarray}
\frac{a_{\ell m}^{\textrm{\scriptsize T}}}{T} &=&
\frac{-3H_{0}^{2}\Omega_{m}}{(2\pi)^{3}c^{3}} \int{\mathrm{d}\Omega \;
Y_{\ell m}^{*}(\mathbf{\hat{n}})\! \int{\!\mathrm{d}r \;
a\frac{\mathrm{d}}{\mathrm{d}t}\left[\frac{g}{a}\right]}} \nonumber \\
&& \times \int{\frac{\mathrm{d}^{3}\mathbf{k}}{k^{2}} \;
\delta_{k}e^{i\mathbf{k}\cdot\mathbf{\hat{n}}r} } \nonumber \\ &=&
\frac{-3H_{0}^{2}\Omega_{m}\mathrm{i}^{\ell}}{2\pi^{2}c^{3}}
\int{\mathrm{d}r \;
a\frac{\mathrm{d}}{\mathrm{d}t}\left[\frac{g}{a}\right]} \nonumber \\
&& \times \int{\frac{\mathrm{d}^{3}\mathbf{k}}{k^{2}} \;
\delta_{k}j_{\ell}(kr)Y_{\ell m}^{*}(\mathbf{\hat{k}})},
\end{eqnarray}
where $\delta$ denotes the matter overdensity field today,
$\mathbf{\hat{k}}$ and $\mathbf{\hat{n}}$ are unit vectors and the
exponential has been re-expressed in terms of spherical Bessel
functions $j_{\ell}$ and spherical harmonics $Y_{\ell m}$.

Within each redshift slice we calculate the projected galaxy
overdensity $\delta_{\mathrm{proj}}$, which is related to the
three-dimensional galaxy overdensity via
\[
\delta_{\textrm{\scriptsize proj}} =
\int{\frac{\bar{n}(r)r^{2}}{\int{\bar{n}(r')r'^{2}}\;\mathrm{d}r'}\delta(r)}\;\mathrm{d}r
\equiv \int{\Theta(r)\delta(r)}\;\mathrm{d}r,
\]
where $\bar{n}(r)$ is the background galaxy number density and $r(z)$
is comoving distance; thus $\Theta(r)$ is the effective radial
weight applied to the density field.

Assuming a linear bias relation to relate galaxy and matter
overdensities $\delta_{g} = b\,\delta_{m}$, the galaxy expansion
coefficients are:
\[
a_{\ell m}^{\textrm{\scriptsize g}} = \frac{b
\,i^{\ell}}{2\pi^{2}}\int{\mathrm{d}r \;
g\,\Theta(r)\int{\mathrm{d}^{3}\mathbf{k} \;
\delta_{k}j_{\ell}(kr)Y_{\ell m}(\mathbf{\hat{k}})}},
\]
where once again the matter overdensity $\delta$ is evaluated today
and $\mathbf{\hat{k}}$ is a unit vector.

Now the cross correlation can be written as
\begin{eqnarray}
C_{\textrm{\scriptsize gT}}(\ell) & = & \frac{-6 b
H_{0}^{2}\Omega_{m}T}{\pi c^{3}} \int{\mathrm{d}k \; P(k)}\int
\!{\mathrm{d}r' \; g\,\Theta(r') j_{\ell}(kr')} \nonumber \\ && \times
\! \int \!{\mathrm{d}r \; g\,H(f-1)} j_{\ell}(kr), \label{eq:cgt}
\end{eqnarray}
where $f=\mathrm{d}\ln g/\mathrm{d}\ln a$.

Limber's approximation can be used to eliminate the spherical Bessel
functions by using the small angle limit
\[
\lim_{\ell \rightarrow \infty} j_{\ell}(x) =
\sqrt{\frac{\pi}{2\ell+1}}\delta^{\textrm{\scriptsize
    K}} \, \left(\ell+\frac{1}{2}-x\right). 
\]
Applying this to equation (\ref{eq:cgt}) gives:
\begin{eqnarray}
C_{\textrm{\scriptsize gT}}(\ell) & = & \frac{-3 b
H_{0}^{2}\Omega_{m}T}{c^{3}(\ell+1/2)^{2}} \int{\mathrm{d}r \;
\Theta(r)Hg^{2}(f-1) } \nonumber \\ && \times
P\left(\frac{\ell+1/2}{r}\right).  \label{eq:cgtlim}
\end{eqnarray}

\begin{figure}
\centering
\rotatebox{270}{
\epsfig{file=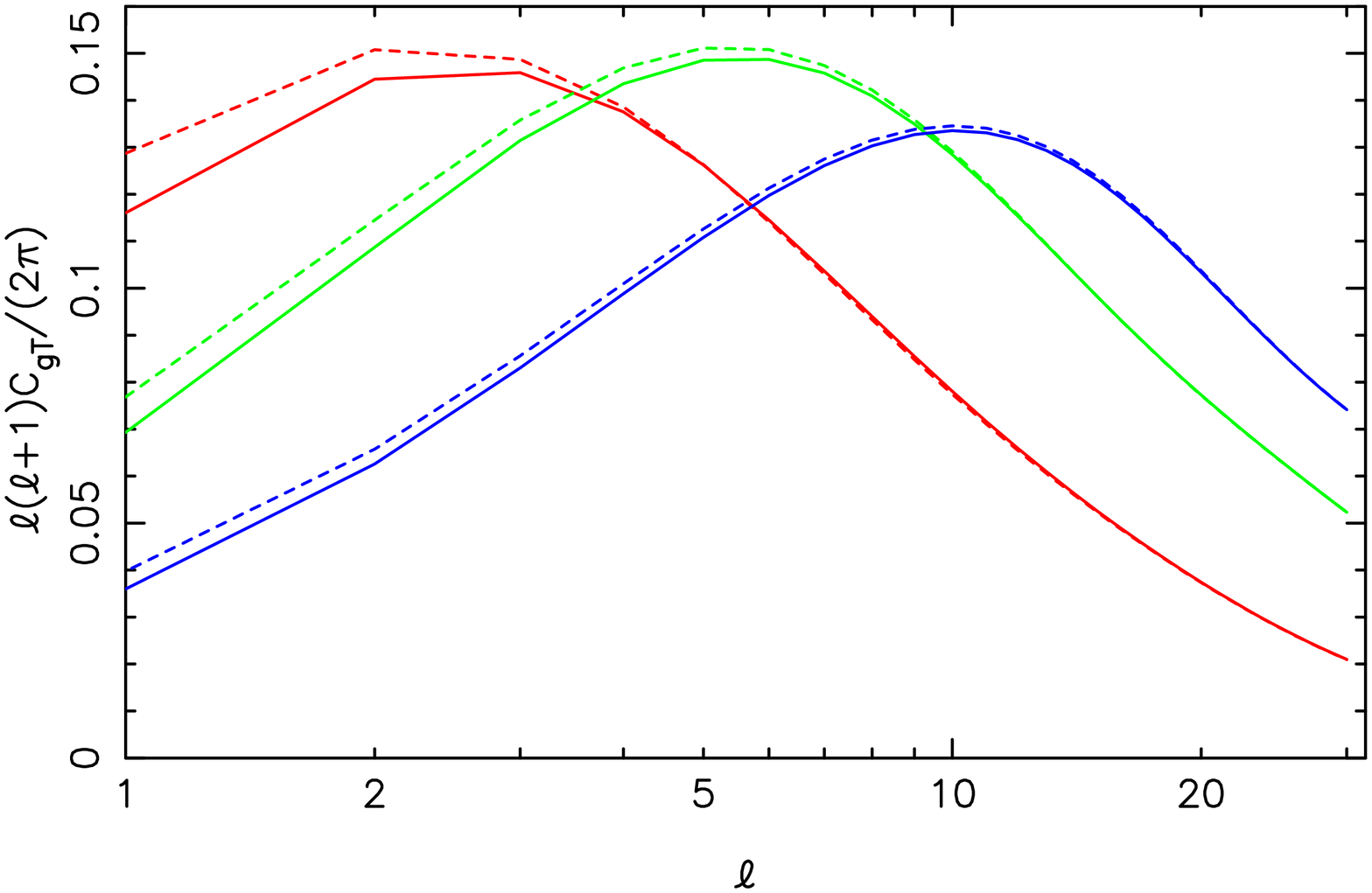,width=6.0cm,height=8.0cm}
}
\caption{The expected cross-correlation signal in $\mu$K for each of
  the 3 redshift slices with the Limber approximation (dashed) and
  exact (solid) lines. The peak of the cross-correlation shifts to
  higher multipoles for higher redshift slices (red: $0<z<0.1$; green: $0.1<z<0.2$; blue: $0.2<z<0.3$). Here a constant bias
  $b=1.4$ has been used. \label{fig:cgts}}
\end{figure}

Fig. \ref{fig:cgts} shows the expected cross-correlation for each of
our galaxy redshift shells. The peak of the cross correlation can be
seen to shift to higher multipoles at higher redshifts as the power
spectrum peak shifts to higher multipoles (a given physical scale will
subtend a smaller angle when viewed from further away, i.e. will
correspond to a higher multipole). The Limber approximation is valid
here for $\ell \gtrsim 10$.

Following an identical method, the galaxy auto-correlation is given by
\[
C_{\textrm{\scriptsize gg}}(\ell) = \frac{2b^{2}}{\pi}\int{\mathrm{d}k
\; k^{2}P(k) \left|\int {\mathrm{d}r\;
g\Theta(r)j_{\ell}(kr)}\right|^{2}}, \label{eq:Cgg}
\]
or on applying the Limber approximation \citep{Kaiser_1992},
\[
C_{\textrm{\scriptsize gg}}(\ell) = b^{2}\int{\mathrm{d}r\;
\frac{\Theta^{2}}{r^{2}}g^{2}P\left(\frac{\ell +
1/2}{r}\right)}. \label{eq:Cgg_lim}
\]

\subsection{Galaxy bias \label{sec:bias}}

The galaxy bias is expected to change with $z$: the furthest redshift
shell is populated only by the brightest, most massive galaxies, which
are known to be more strongly clustered
\citep{Park_1994,Loveday_1995}. In order to determine the bias in each
redshift slice, we use the galaxy angular power spectra. The form of
the angular power spectrum in each slice can be predicted given a
three-dimensional power spectrum using equations (\ref{eq:Cgg}) and
(\ref{eq:Cgg_lim}). The effective bias of the predicted spectrum can
then be adjusted to match observations. In linear theory, bias cannot
be determined independently of $\sigma_{8}$ which also acts to
renormalize the power spectrum, $C_{\ell} \propto
(b\sigma_{8})^{2}$. In what follows, we therefore fix the value of
$\sigma_{8}$ at 0.75 and fit only for $b$.

\begin{figure*}
\centering
\begin{minipage}[l]{5.0cm}
\centering
\rotatebox{270}{
\epsfig{file=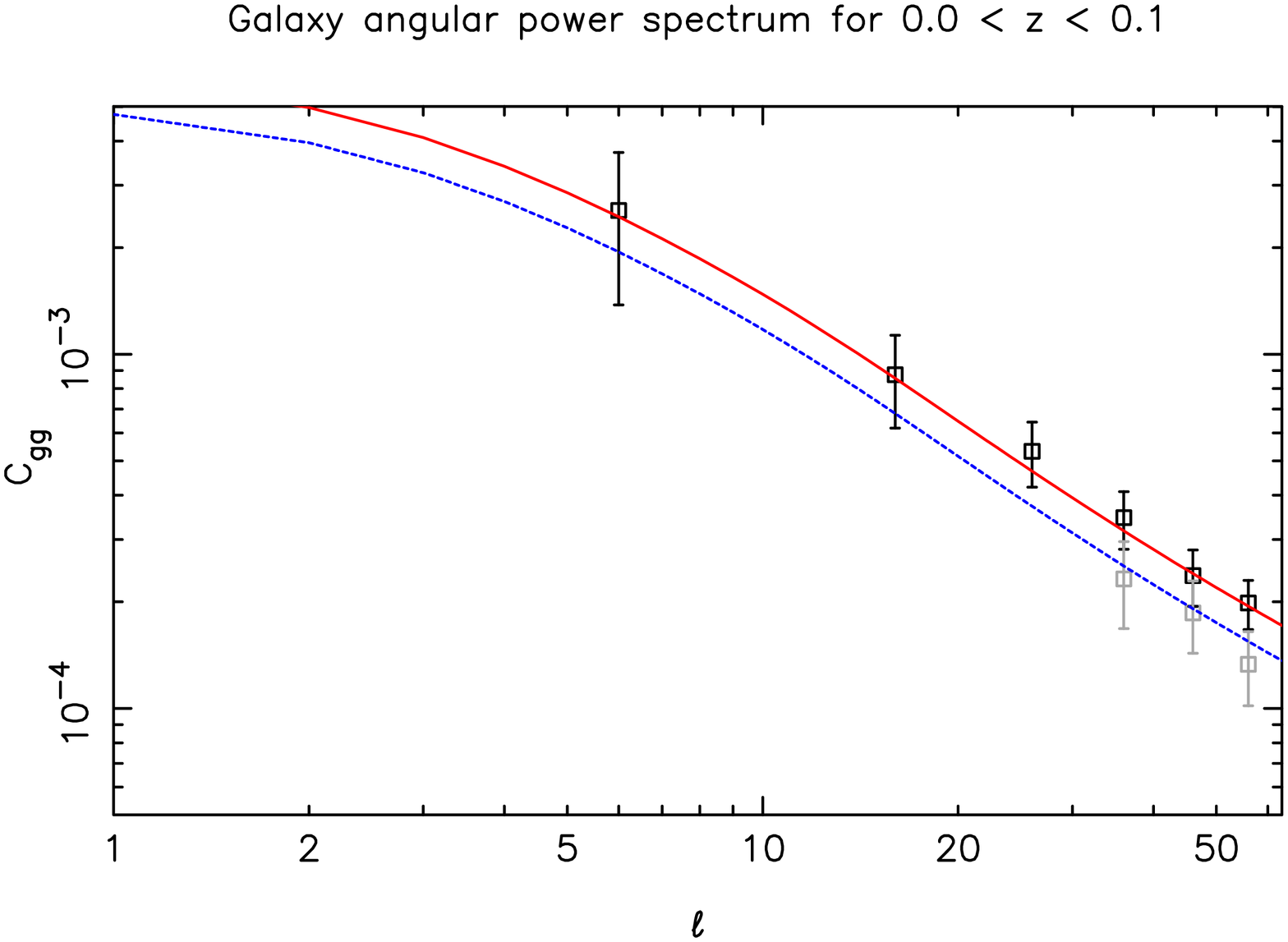,width=6.5cm,height=6.3cm}
}
\end{minipage}
\hspace{0.7cm}
\begin{minipage}[r]{5.0cm}
\centering
\rotatebox{270}{
\epsfig{file=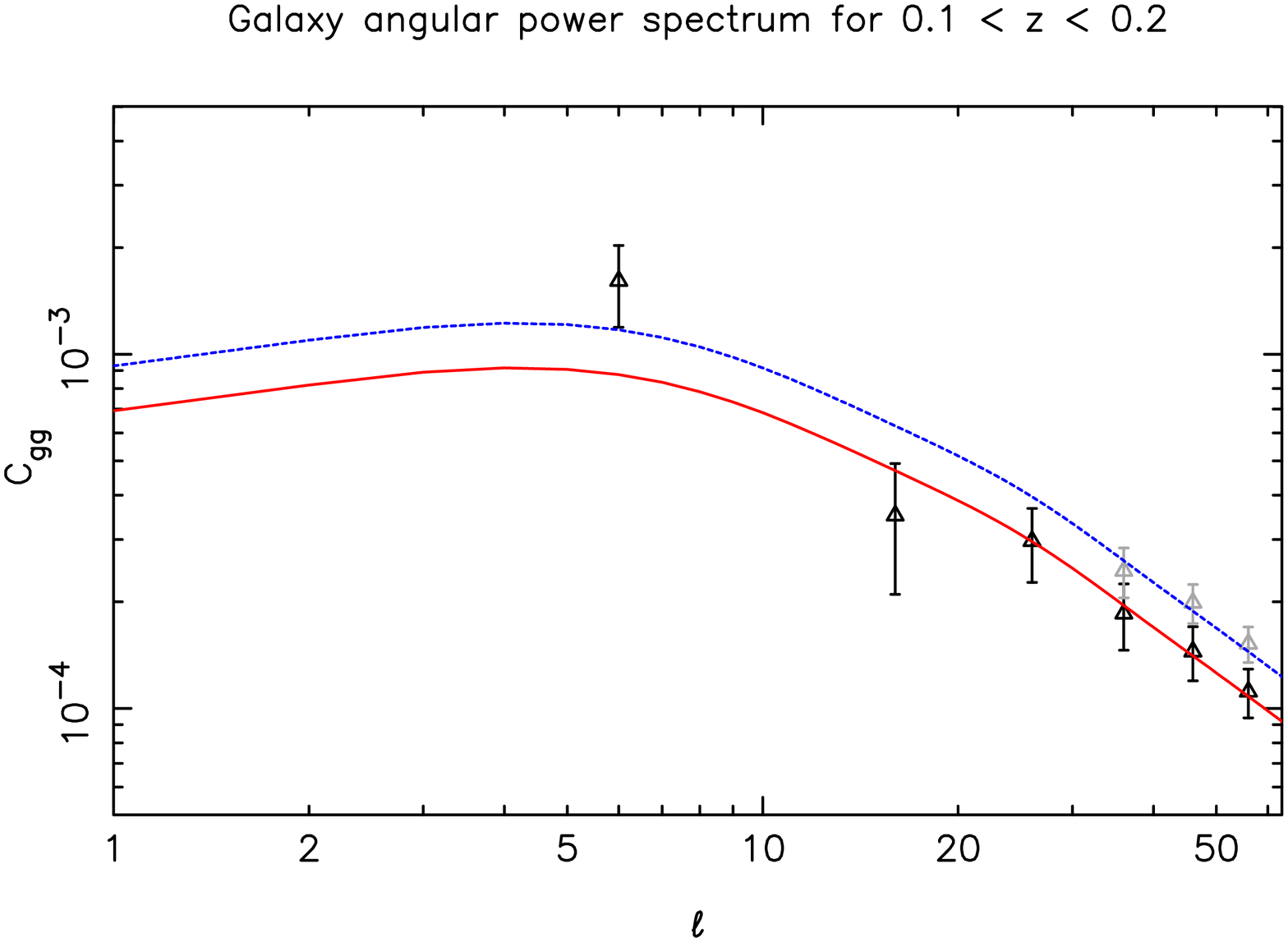,width=6.5cm,height=6.3cm}
}
\end{minipage}
\hspace{0.7cm}
\begin{minipage}[l]{5.0cm}
\centering
\rotatebox{270}{
\epsfig{file=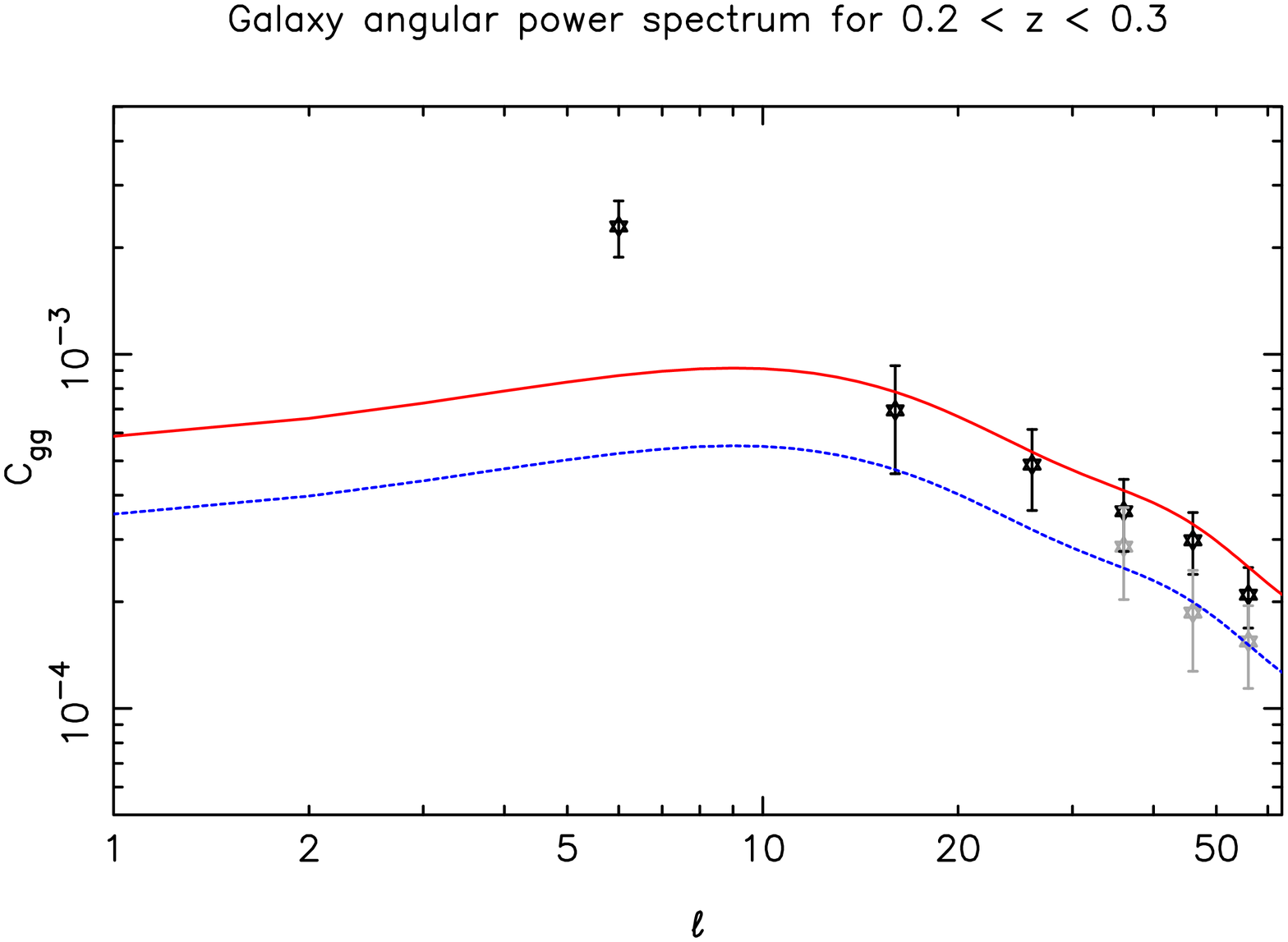,width=6.5cm,height=6.3cm}
}
\end{minipage}
\hspace{0.7cm}
\caption{Fits to the measured galaxy angular power spectra in each
  redshift slice used to determine the bias. The bold symbols show the
  photometric data and the grey symbols the spectroscopic data
  (at $\ell>30$ only). Fits to the
  photometric data alone and the spectroscopic data alone are shown by
  the red solid and blue dotted lines respectively. \label{fig:Cgg_fit}}
\end{figure*}

We use CAMB \citep{Lewis_CAMB} to generate non-linear matter power
spectra for our cosmological model (see Section \ref{sec:intro}) and
output the three-dimensional spectra at redshifts at the mid-point of
each redshift shell in question. Although only strictly valid in
linear theory, we evolve the non-linear power spectrum for each slice
according to the growth function; this is thought to be a good
approximation to the non-linear power spectrum on the linear and
quasi-linear scales where we will fit for the bias. We have two
sources of data that can be used to measure the angular power in each
redshift slice: the photometric redshift data and the subset of this
in the SDSS region where complete spectroscopic redshifts exist. Since
the spectroscopic data exist only in a subregion, the angular power
measured in this case will not be accurate on large scales. However,
we should expect reasonable agreement between the photometric angular
power and the spectroscopic angular power on smaller scales, up to
differences due to sample variance and mask. Fig. \ref{fig:Cgg_fit}
shows the shot noise corrected photometric and spectroscopic angular
power spectra binned for $\ell \le 60$ in bins of width $\Delta\ell =
10$ and the best-fitting predicted spectra for each of the redshift
slices.

The fit is made using a
maximum likelihood approach where the error bars are due to cosmic
variance alone and are calculated from the model power spectra. The likelihood is 
\[
{\mathcal L} \propto |\mathbf{\mathsf{M}}|^{-1/2}\exp\{-(\mathbf{d}^{\textrm \scriptsize
    T}\mathbf{\mathsf{M}}^{-1}\mathbf{d})/2\},
\]
assuming the covariance matrix $\mathbf{\mathsf{M}}$ to be diagonal
and therefore neglecting correlations between multipole bins
\citep{Blake_SDSSLRG}. The vector $\mathbf{d}$ is of differences
between model and data, $d_{i} = C_{\mathrm{gg}}^{\textrm{\scriptsize
model}} - C_{\mathrm{gg}}^{\textrm{\scriptsize data}}$ in each bin,
and the diagonal elements of $\mathbf{\mathsf{M}}$ are
\[
\sigma(C_{\mathrm{gg}}) = \sqrt{\frac{2}{f_{\textrm{\scriptsize sky}}(2\ell+1)}}C_{\mathrm{gg}}.
\]
The fit is performed firstly for the
photometric redshift data alone and secondly for the spectroscopic
data alone. The best-fitting values for the bias are given in Table
\ref{table:bias}. Both the spectroscopic and photometric data show an
increase in bias with redshift, which gives us confidence that this effect is real.
The difference in bias between all photometric redshifts and 
SDSS true-$z$ information is small in comparison with this trend;
given the restricted overlap of these samples and the correlated errors
shown in Fig. \ref{fig:Cgg_fit}, we believe that it is insignificant
within cosmic variance.

\begin{table}
\begin{tabular}{cccc}\hline
&$0.0<z<0.1$&$0.1<z<0.2$&$0.2<z<0.3$ \\ \hline
photo-$z$ only & 1.22 & 1.65 & 2.86 \\
SDSS only & 1.03 & 1.90 & 2.22 \\ \hline
\end{tabular}
\caption{Best fit values for the bias calculated using the photometric
redshift data only and the photometric data replaced by SDSS data
using regions where the latter information is
available. \label{table:bias}}
\end{table}


\section{Cross-Correlation Analysis}

\label{sec:cross-corr}

In this work, we have two hypotheses: a null hypothesis
$C_{\textrm{\scriptsize gT}} = 0$, and a hypothesis where the
cross-correlation is as expected in a $\Lambda$CDM universe and is
described by equations (\ref{eq:cgt}) and (\ref{eq:cgtlim}). We assume
that the cosmological model is given; there are thus no adjustable
parameters to consider, and the hypotheses can be compared by means of
a simple likelihood ratio.

The cross-correlation for a pair of galaxy and temperature maps is calculated using
\[
C_{\textrm{\scriptsize gT}} = \frac{1}{(2\ell+1)}
\sum_{m=-\ell}^{\ell} \frac{a_{\ell
m}^{\mathrm{g}}}{\sqrt{f_{\mathrm{sky}}^{\mathrm{g}}}} \frac{a_{\ell
m}^{\mathrm{T*}}}{\sqrt{f_{\mathrm{sky}}^{\mathrm{T}}}},
\label{eq:cgtcomp}
\]
where the $f_{\textrm{\scriptsize sky}}$ factors attempt to compensate
for the loss of sky coverage due to the masks. In detail, masking also
introduces correlations between multipoles, which we take into account
by using a full covariance matrix to calculate the $\chi^{2}$
statistic; see Section \ref{sec:covmat}. The cross-correlation data
are binned to improve the signal-to-noise; we use 5 logarithmically
spaced bins for $3 \le \ell \le 30$ and ignore $\ell=2$ for
consistency with R07. From our 3 redshift slices with 5 multipole bins
per slice, we have 15 data points in total to analyse for each WMAP
band.

\begin{figure*}
\centering
\begin{minipage}[l]{5.0cm}
\centering
\rotatebox{270}{
\epsfig{file=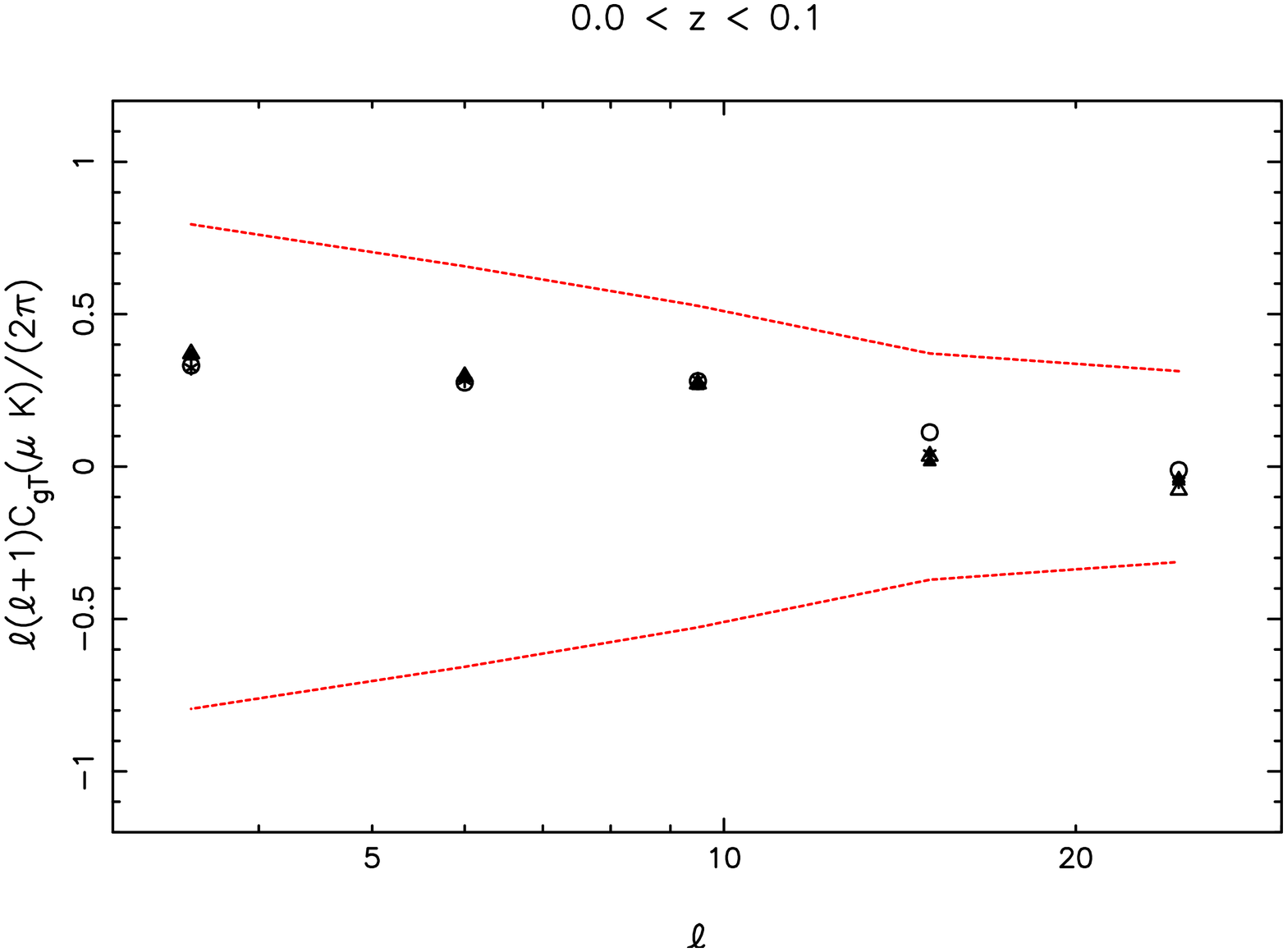,width=6.0cm,height=6.0cm}
}
\end{minipage}
\hspace{0.7cm}
\begin{minipage}[r]{5.0cm}
\centering
\rotatebox{270}{
\epsfig{file=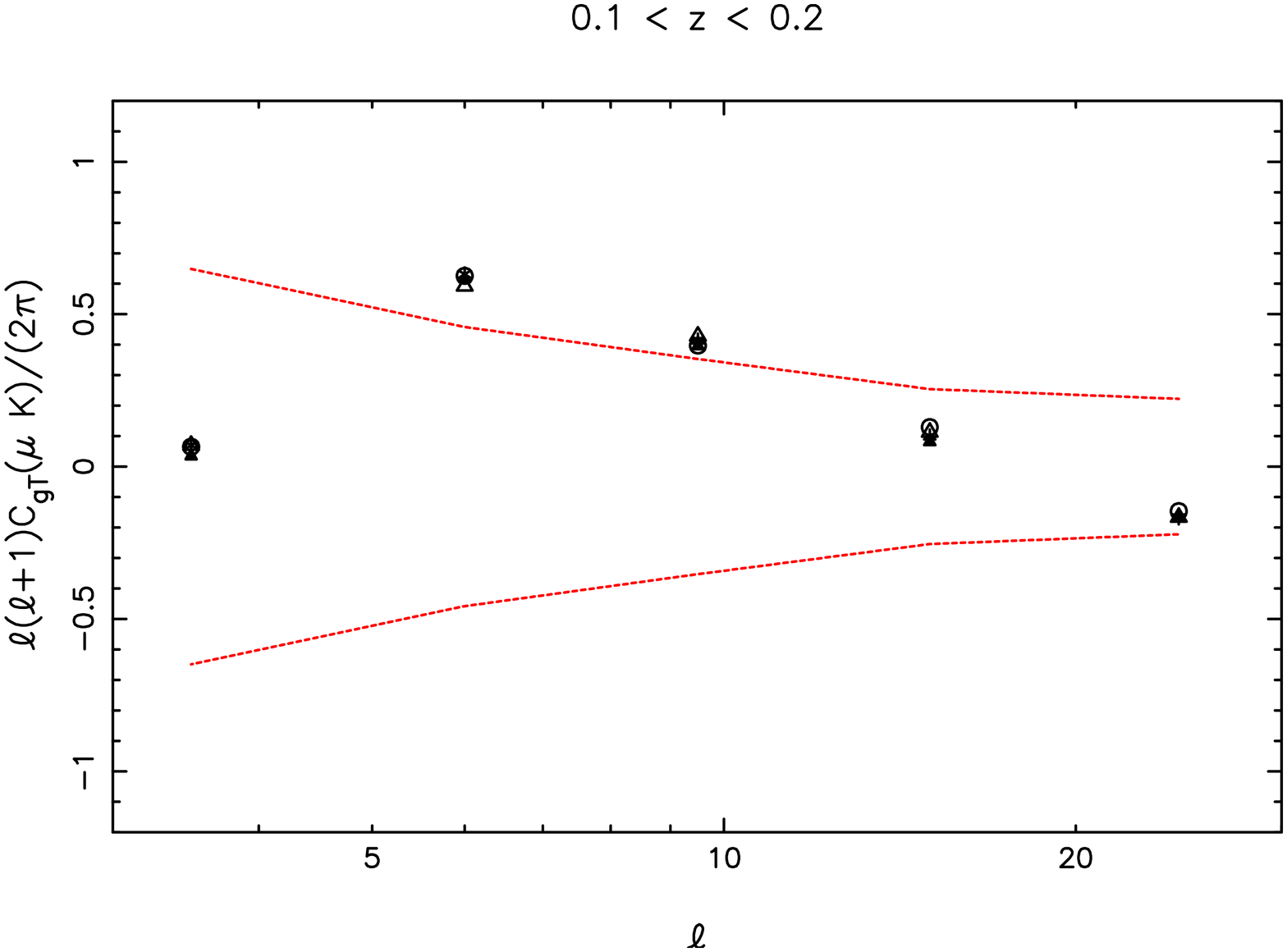,width=6.0cm,height=6.0cm}
}
\end{minipage}
\hspace{0.7cm}
\begin{minipage}[l]{5.0cm}
\centering
\rotatebox{270}{
\epsfig{file=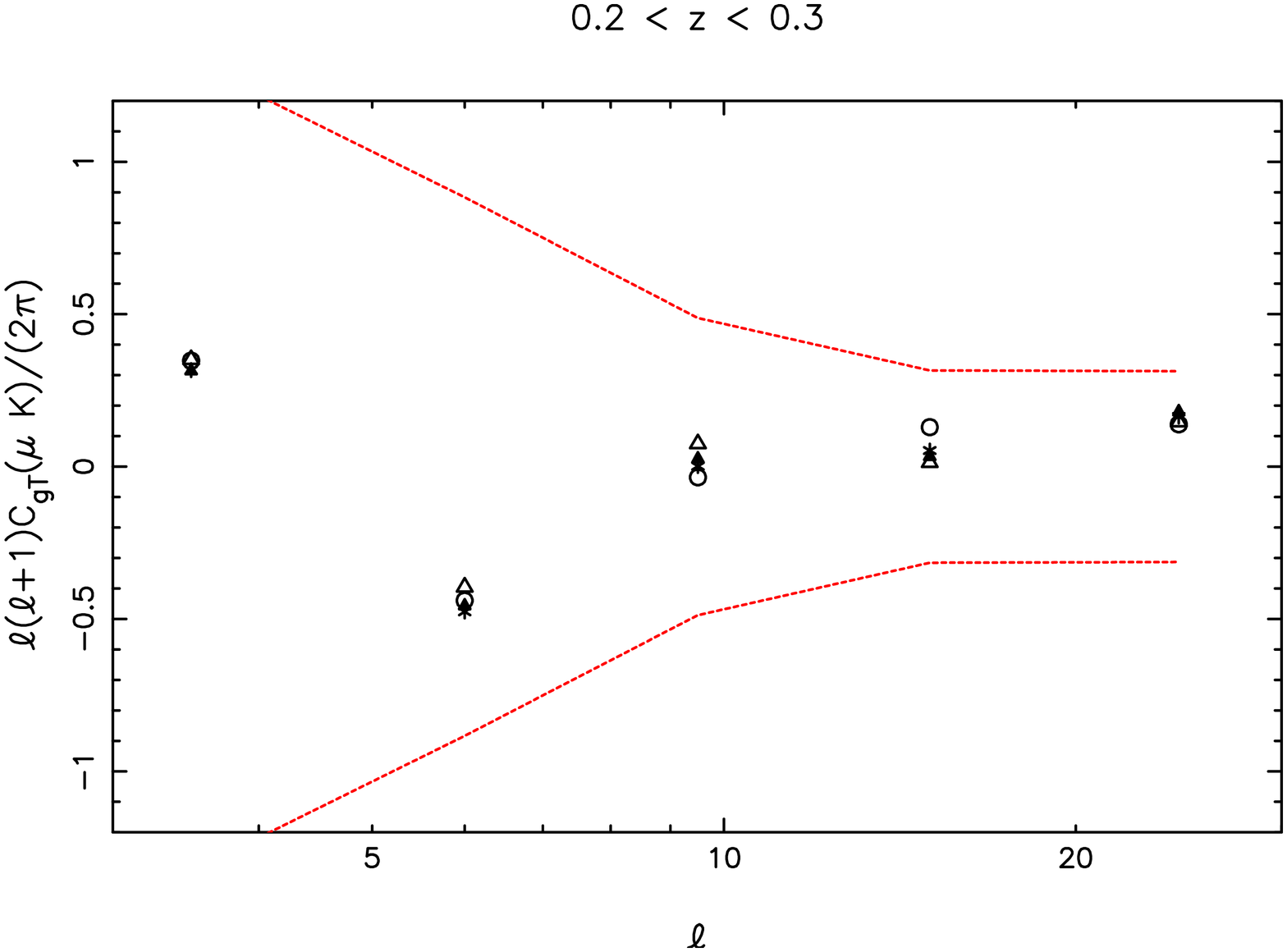,width=6.0cm,height=6.0cm}
}
\end{minipage}
\hspace{0.7cm}
\caption{The cross-correlation results for each of the photometric
  redshift slices using all the WMAP data: ILC (open triangle), $Q$
  (open circle), $V$ (star) and $W$ (filled triangle). We see from the
  $1\sigma$ error bars around the null hypothesis (shown as lines)
  that the data are consistent with no ISW
  effect. \label{fig:cgtdata}}
\end{figure*}

The cross-correlations measured in each redshift slice for each of the
WMAP bands are shown in Fig. \ref{fig:cgtdata}. The results are the
same for each of the WMAP temperature maps considered, which is
certainly consistent with the achromatic nature of the ISW
effect. However, the plots also show the rms dispersion in
$C_{\textrm{\scriptsize gT}}$ under the null hypothesis of no ISW
effect, from which it is clear that no signal is significantly detected. We quantify
this by evaluating the $\chi^{2}$ statistic for each hypothesis using
\[
\chi^{2} = \sum_{i,j} d_{i}(\mathsf{C}^{-1})_{ij}d_{j} \label{eq:chisq},
\]
where ${\mathsf{C}}$ is the covariance matrix computed from
simulations (see Section \ref{sec:covmat}). The $d_{i}$ are the values
of $\smash{(C_{\mathrm{gT}}^{\mathrm{data}} -
C_{\mathrm{gT}}^{\mathrm{hyp}})}$ where
$\smash{C_{\mathrm{gT}}^{\mathrm{hyp}}}$ is either zero or given by
equations (\ref{eq:cgt}) and (\ref{eq:cgtlim}) for the fiducial
hypothesis; $C_{\mathrm{gT}}^{\mathrm{data}}$ are the measured
cross-correlation values and $i$ labels the binned measurements.

The likelihood of each hypothesis is determined using 
\[
\mathcal{L} \propto
    |\mathsf{C}|^{-1/2}\exp\left\{-(\mathbf{d}^{\textrm{\scriptsize
    T}}\mathsf{C}^{-1}\mathbf{d})/2\right\},
\]
with $\mathbf{d}$ the data vector whose components are defined above
and $\mathsf{C}$ the covariance matrix. In practice, the covariance
matrix used in each case is calculated from simulations of the null
hypothesis; this is a reasonable approximation given the subdominant
nature of the ISW effect. Then
\[
-2\ln\left\{\frac{\mathcal{L}_{1}}{\mathcal{L}_{2}}\right\} = \Delta\chi^{2},
\]
where $\mathcal{L}_{1}$ and $\mathcal{L}_{2}$ are the likelihoods for
the two hypotheses.

\subsection{Covariance matrix estimation\label{sec:covmat}}

In order to calculate the covariance matrix for the $C_{\textrm{gT}}$
values, we use 500 simulations each of independent CMB and galaxy
density fields. The power spectrum used for the CMB Gaussian
realizations is the best-fitting theoretical angular power spectrum
found by the WMAP team \citep{Spergel_WMAP3}.

We generate simulations of the galaxy density fields by assuming a
lognormal form for the measured 2MASS density field
\citep{Coles_Jones}. This allows a rapid generation of realistic
density fields, from which mock galaxies are drawn via Poisson
sampling. We now have a set of simulated CMB and 2MASS galaxy maps
with the same shot noise properties as the data.

\begin{figure*}
\centering
\begin{minipage}[l]{5.0cm}
\centering
\rotatebox{270}{
\epsfig{file=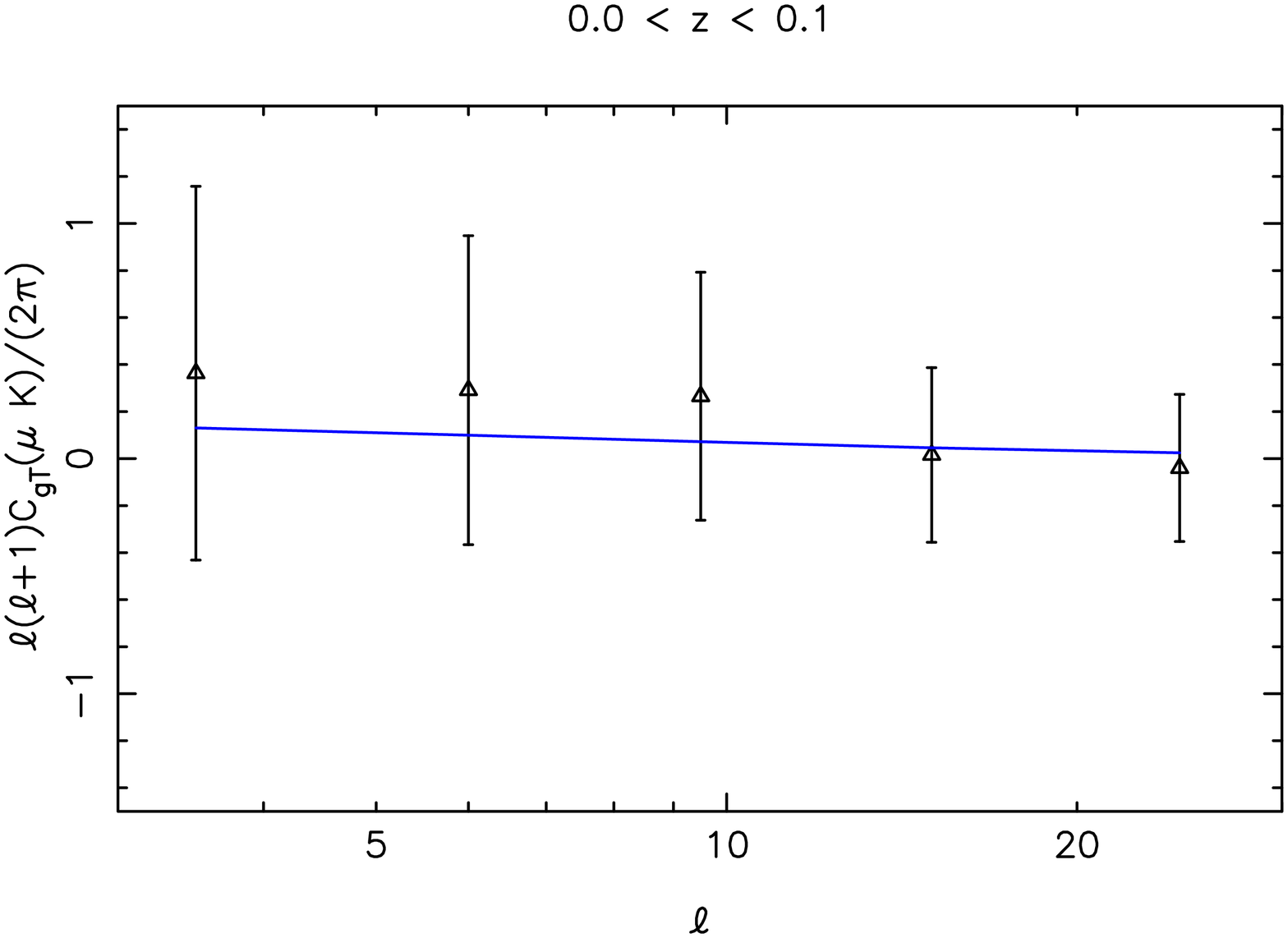,width=6.0cm,height=6.0cm}
}
\end{minipage}
\hspace{0.7cm}
\begin{minipage}[r]{5.0cm}
\centering
\rotatebox{270}{
\epsfig{file=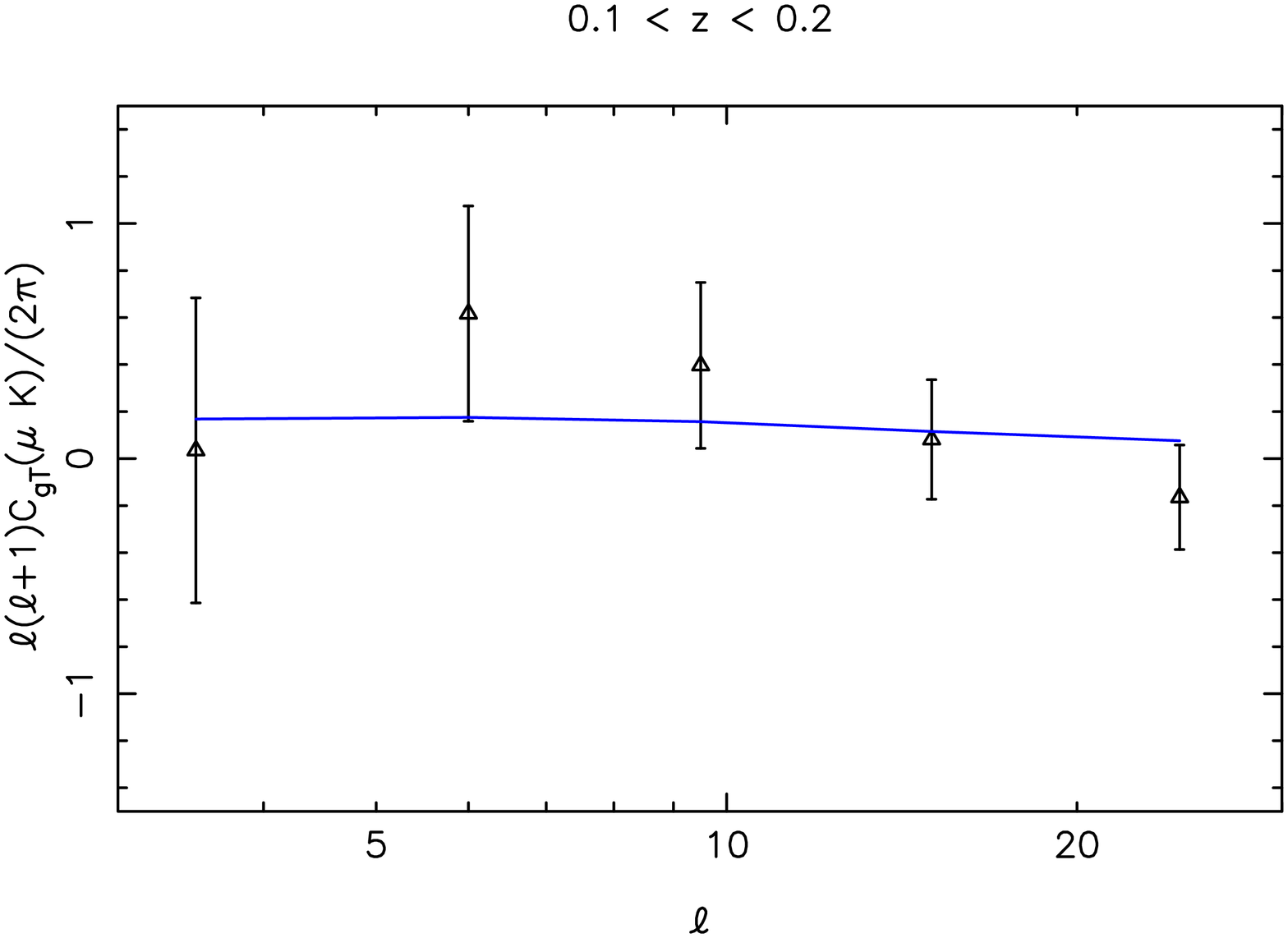,width=6.0cm,height=6.0cm}
}
\end{minipage}
\hspace{0.7cm}
\begin{minipage}[l]{5.0cm}
\centering
\rotatebox{270}{
\epsfig{file=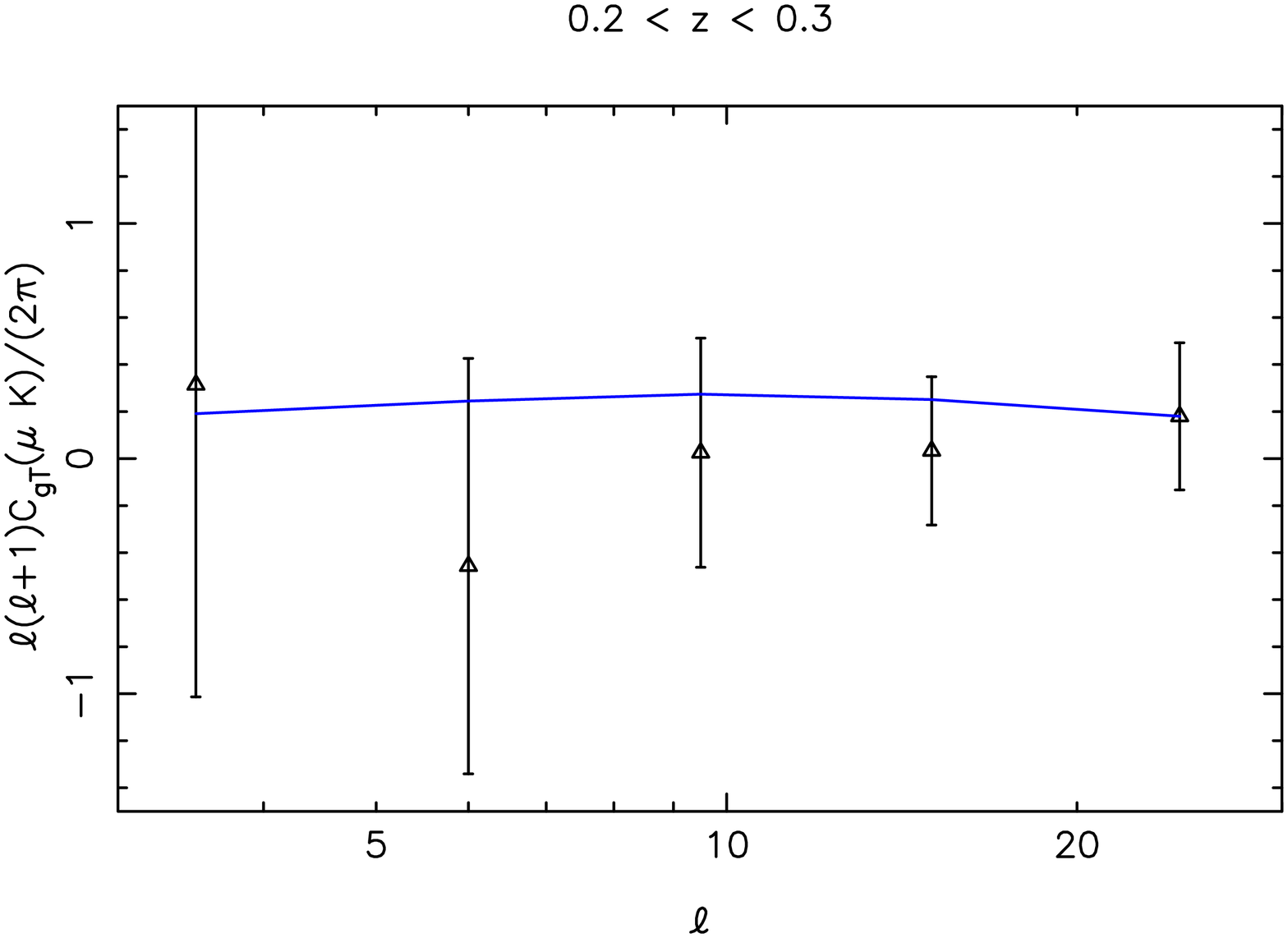,width=6.0cm,height=6.0cm}
}
\end{minipage}
\hspace{0.7cm}
\caption{The cross-correlation data for the $V$-band CMB data in the
  different redshift slices. Error bars are calculated from simulated
  CMB skies and 2MASS galaxy maps. The expected ISW signal in a $\Lambda$CDM
  universe is also shown calculated for bias values measured from the
  photometric data (see Table \ref{table:bias}).\label{fig:errorbars}}
\end{figure*}

The covariance matrix for the null hypothesis is estimated from the simulated data
\[
\mathsf{C}_{ij} = \left\langle(d_{i}-\left\langle
d_{i}\right\rangle)(d_{j}-\left\langle d_{j}\right\rangle)
\right\rangle,
\]
where the data vector $\mathbf{d}$ is defined as in equation
(\ref{eq:chisq}). Using only simulations of the CMB correlated with
the observed galaxy maps leads to an underestimation of error bars at
the $10\%$ level on all scales, in agreement with \citet{Cabre_2007}.
The error bars calculated from the diagonal elements of the covariance
matrix are plotted in Fig. \ref{fig:errorbars}, and contrasted with
the ISW signal expected in our default $\Lambda$CDM universe.

\subsection{Hypothesis testing\label{sec:hyp_test}}

The $\chi^{2}$ values for each hypothesis are given in Tables
\ref{table:chisq}, \ref{table:chisqsdss} and \ref{table:chisqb1.4}
considering bias values computed from photometric data only, SDSS data
only and for $b=1.4$ consistent with R07 respectively. Each table has
data for each redshift slice and for an analysis of all three slices
together. The results depend to some extent on the values adopted for
the bias in each slice. This does not affect the covariance matrix:
values for the bias only affect the final $\chi^{2}$ results through
an alteration of the expected $C_{\textrm{\scriptsize gT}}$ signal.

\begin{table*}
\begin{tabular}{cc|ccc|ccc|ccc|cccc} \hline
& \hspace{0.5cm} & \multicolumn{2}{c|}{$0.0<z<0.1$} & \hspace{0.5cm} &
\multicolumn{2}{c|}{$0.1<z<0.2$} & \hspace{0.5cm} &
\multicolumn{2}{c|}{$0.2<z<0.3$} & \hspace{0.5cm} &
\multicolumn{3}{c|}{$0.0<z<0.3$} \\ \hline WMAP Band & \hspace{0.5cm}
& $\chi^{2}_{\textrm{\scriptsize null}}$
&$\chi^{2}_{\textrm{\scriptsize fid}}$ & \hspace{0.5cm} &
$\chi^{2}_{\textrm{\scriptsize null}}$ &$\chi^{2}_{\textrm{\scriptsize
fid}}$ & \hspace{0.5cm} & $\chi^{2}_{\textrm{\scriptsize null}}$
&$\chi^{2}_{\textrm{\scriptsize fid}}$ & \hspace{0.5cm} &
$\chi^{2}_{\textrm{\scriptsize null}}$ &$\chi^{2}_{\textrm{\scriptsize
fid}}$ & $\Delta\chi^{2}$ \\ \hline ILC &\hspace{0.5cm} & 0.70 & 0.41
& \hspace{0.5cm}& 3.8 & 2.6 &\hspace{0.5cm} & 0.52 & 1.3
&\hspace{0.5cm} & 5.0 & 4.3 & 0.7 & \\ $Q$ &\hspace{0.5cm} & 0.68 &
0.32 & \hspace{0.5cm}& 3.7 & 2.5 &\hspace{0.5cm} & 0.68 & 1.2
&\hspace{0.5cm} & 5.1 & 3.9 & 1.2 & \\ $V$ &\hspace{0.5cm} & 0.63 &
0.33 & \hspace{0.5cm}& 3.9 & 2.8 &\hspace{0.5cm} & 0.65 & 1.4
&\hspace{0.5cm} & 5.2 & 4.4 & 0.8 & \\ $W$ &\hspace{0.5cm} & 0.63 &
0.33 & \hspace{0.5cm}& 3.6 & 2.6 &\hspace{0.5cm} & 0.67 & 1.3
&\hspace{0.5cm} & 4.9 & 4.3 & 0.6 & \\ \hline
\end{tabular}
\caption{$\chi^{2}$ values for the two hypotheses for each WMAP band
  analysed. The first three sections of the table show the $\chi^{2}$
  values for the three redshift slices separately, the final section
  of the table shows $\chi^{2}$ and $\Delta\chi^{2}$ values for the
  entire data set. We see that the first two redshift slices prefer a
  $\Lambda$CDM ISW signal whilst the third prefers the null
  hypothesis. The whole dataset prefers a $\Lambda$CDM ISW signal. The
  significance of this result is discussed in
  \ref{sec:hyp_test}. \label{table:chisq}}
\end{table*}

\begin{table*}
\begin{tabular}{cc|ccc|ccc|ccc|cccc} \hline
& \hspace{0.5cm} & \multicolumn{2}{c|}{$0.0<z<0.1$} & \hspace{0.5cm} &
\multicolumn{2}{c|}{$0.1<z<0.2$} & \hspace{0.5cm} &
\multicolumn{2}{c|}{$0.2<z<0.3$} & \hspace{0.5cm} &
\multicolumn{3}{c|}{$0.0<z<0.3$} \\ \hline WMAP Band & \hspace{0.5cm}
& $\chi^{2}_{\textrm{\scriptsize null}}$
&$\chi^{2}_{\textrm{\scriptsize fid}}$ & \hspace{0.5cm} &
$\chi^{2}_{\textrm{\scriptsize null}}$ &$\chi^{2}_{\textrm{\scriptsize
fid}}$ & \hspace{0.5cm} & $\chi^{2}_{\textrm{\scriptsize null}}$
&$\chi^{2}_{\textrm{\scriptsize fid}}$ & \hspace{0.5cm} &
$\chi^{2}_{\textrm{\scriptsize null}}$ &$\chi^{2}_{\textrm{\scriptsize
fid}}$ & $\Delta\chi^{2}$ \\ \hline ILC &\hspace{0.5cm} & 0.70 & 0.44
& \hspace{0.5cm}& 3.8 & 2.6 &\hspace{0.5cm} & 0.52 & 0.86
&\hspace{0.5cm} & 5.0 & 3.9 & 1.1 & \\ $Q$ &\hspace{0.5cm} & 0.68 &
0.36 & \hspace{0.5cm}& 3.7 & 2.4 &\hspace{0.5cm} & 0.68 & 0.82
&\hspace{0.5cm} & 5.1 & 3.5 & 1.6 & \\ $V$ &\hspace{0.5cm} & 0.63 &
0.37 & \hspace{0.5cm}& 3.9 & 2.7 &\hspace{0.5cm} & 0.65 & 0.96
&\hspace{0.5cm} & 5.2 & 4.0 & 1.2 & \\ $W$ &\hspace{0.5cm} & 0.63 &
0.37 & \hspace{0.5cm}& 3.6 & 2.6 &\hspace{0.5cm} & 0.67 & 0.96
&\hspace{0.5cm} & 4.9 & 3.9 & 1.0 & \\ \hline
\end{tabular}
\caption{As above except that bias values in the three redshift slices
  are calculated using spectroscopic measurements of the angular power
  spectrum (see Section \ref{sec:hyp_test}).\label{table:chisqsdss}}
\end{table*}

\begin{table*}
\begin{tabular}{cc|ccc|ccc|ccc|cccc} \hline
& \hspace{0.5cm} & \multicolumn{2}{c|}{$0.0<z<0.1$} & \hspace{0.5cm} &
\multicolumn{2}{c|}{$0.1<z<0.2$} & \hspace{0.5cm} &
\multicolumn{2}{c|}{$0.2<z<0.3$} & \hspace{0.5cm} &
\multicolumn{3}{c|}{$0.0<z<0.3$} \\ \hline WMAP Band & \hspace{0.5cm}
& $\chi^{2}_{\textrm{\scriptsize null}}$
&$\chi^{2}_{\textrm{\scriptsize fid}}$ & \hspace{0.5cm} &
$\chi^{2}_{\textrm{\scriptsize null}}$ &$\chi^{2}_{\textrm{\scriptsize
fid}}$ & \hspace{0.5cm} & $\chi^{2}_{\textrm{\scriptsize null}}$
&$\chi^{2}_{\textrm{\scriptsize fid}}$ & \hspace{0.5cm} &
$\chi^{2}_{\textrm{\scriptsize null}}$ &$\chi^{2}_{\textrm{\scriptsize
fid}}$ & $\Delta\chi^{2}$ \\ \hline ILC &\hspace{0.5cm} & 0.70 & 0.38
& \hspace{0.5cm}& 3.8 & 2.7 &\hspace{0.5cm} & 0.52 & 0.55
&\hspace{0.5cm} & 5.0 & 3.6 & 1.4 & \\ $Q$ &\hspace{0.5cm} & 0.68 &
0.28 & \hspace{0.5cm}& 3.7 & 2.6 &\hspace{0.5cm} & 0.68 & 0.58
&\hspace{0.5cm} & 5.1 & 3.4 & 1.7 & \\ $V$ &\hspace{0.5cm} & 0.63 &
0.30 & \hspace{0.5cm}& 3.9 & 2.8 &\hspace{0.5cm} & 0.65 & 0.66
&\hspace{0.5cm} & 5.2 & 3.6 & 1.6 & \\ $W$ &\hspace{0.5cm} & 0.63 &
0.30 & \hspace{0.5cm}& 3.6 & 2.7 &\hspace{0.5cm} & 0.67 & 0.67
&\hspace{0.5cm} & 4.9 & 3.8 & 1.1 & \\ \hline
\end{tabular}
\caption{As above but assuming a constant bias $b=1.4$ in all three
  redshift slices as in R07. \label{table:chisqb1.4}}
\end{table*}

In general, the $\chi^{2}$ results are low for the number of degrees
of freedom; this was also noted by R07. We would expect values around
$\chi^{2} = 5 \pm\sqrt{10}$ for the individual redshift slices and
$\chi^{2} = 15 \pm\sqrt{30}$ for all slices together, but only the
second redshift slice gives $\chi^{2}$ values that fall within these
expected ranges. An analytic estimation of the error bars
\citep{Afshordi},
\[
\sigma^{2}(C_{\textrm{\scriptsize gT}}) =
\frac{1}{f_{sky}(2\ell+1)}(C_{\textrm{\scriptsize gT}}^{2} +
C_{\textrm{\scriptsize gg}}C_{\textrm{\scriptsize TT}}),
\]
suggests errors that are even larger than those used here except for
the lowest $\ell$ bin in the third redshift slice, so over-estimation
of our error bars seems unlikely.

For the total analysis using the biases calculated from the
photometric data alone, $\Delta\chi^{2} =
\chi^{2}_{\textrm{\scriptsize null}}-\chi^{2}_{\textrm{\scriptsize
fid}} \simeq 0.825$, which translates to a likelihood ratio of
\[
\frac{\mathcal{L}(\textrm{\small ISW}|\textrm{\small
data})}{\mathcal{L}(\textrm{\small No ISW}|\textrm{\small data})}
\simeq e^{0.41} \simeq 1.5.
\]

This is a long way from decisive evidence for the ISW effect. Using
the other bias values we prefer the fiducial hypothesis at levels of
$1.8 : 1$ and $2 : 1$ for the SDSS bias and $b=1.4$ respectively. This
latter scenario ($b=1.4$) is used by R07 and we find $\Delta\chi^{2}$
values very similar to their results in this case: $\Delta\chi^{2}
\simeq 1.5$ against their $\Delta\chi^{2} \simeq 1.6$. The differences
between the analyses are our use of photometric redshifts, our
slightly different magnitude cut of $12 < K < 13.8$ rather than $12 <
K < 14$ and our use of simulations of the galaxy data as well as the
CMB data to compute errors. As a result of simulating both CMB and
galaxy data, we would expect our $\chi^{2}$ values to be smaller, as
indeed is the case. Thus the main result of this paper is that photometric
redshift information does not increase the sensitivity of the 2MASS
data to the ISW effect.


\section{Statistical power of current and future ISW experiments}

\label{sec:stat_power}

Given the lack of a significant ISW detection from 2MASS, one is led to ask whether this is
as expected, or whether we have been unlucky. We can address this by
using our simulation apparatus to generate the distribution of
$\Delta\chi^{2}$ values we could expect if the no-ISW null hypothesis were
true
\[
P(\Delta\chi^{2} = x\mid H_{0} \textrm{ true}).
\]
Here, the statistic $\Delta\chi^{2}$ itself is computed from the data
in comparison with both $H_{0}$ the null hypothesis, and $H_{1}$ the
alternative hypothesis \emph{but for data generated where $H_{0}$ is
true}. This distribution therefore shows how frequently a given
threshold in $\Delta\chi^{2}$ would reject the null hypothesis when it
is in fact true (the probability of making a Type I error).

In a similar manner, we can obtain a distribution of $\Delta\chi^{2}$
values under the assumption that the alternative hypothesis
($\Lambda$CDM) is true. For each pair of simulated CMB and galaxy
maps, we add to the CMB map the expected ISW effect for the galaxy map
in question. The expected ISW signal can be computed by using equation
(\ref{eq:isw}) together with Poisson's equation to find:
\begin{eqnarray}
\frac{\Delta T_{\ell m}}{T} \!\!\!&=&\!\!\! -2\int
\frac{d}{dt}\left[\frac{g(a)}{a}\right]{\frac{a^{2}\Phi_{\ell
m}(a)}{g(a)}} \, \frac{dr}{c^{3}} \nonumber \\ \!\!\!&\simeq&\!\!\!
\frac{-2}{c^{3}}H(\bar{a})
\left(\frac{dg}{da}(\bar{a})-\frac{g(\bar{a})}{\bar{a}}\right)\frac{\bar{a}^{2}}{g(\bar{a})}
\int{\Phi_{\ell m}} \, dr \nonumber \\ \!\!\!&\simeq&\!\!\!
\frac{3H_{0}^{2}\Omega_{m}}{\ell(\ell+1)c^{3}}\left(1 -
\bar{a}\frac{g'(\bar{a})}{g(\bar{a})}\right)r^{2}(\bar{a})H(\bar{a})
\frac{\delta_{\ell m}}{b}\Delta r, \label{eq:iswmap}
\end{eqnarray}
where $\delta$ is the projected galaxy density field in the redshift
shell under consideration and $\bar{a} = (1+\bar{z})^{-1}$, where
$\bar{z}$ is the redshift at the midpoint of the shell.

\begin{figure*}
\centering
\begin{minipage}[l]{7.0cm}
\centering
\rotatebox{270}{
\epsfig{file=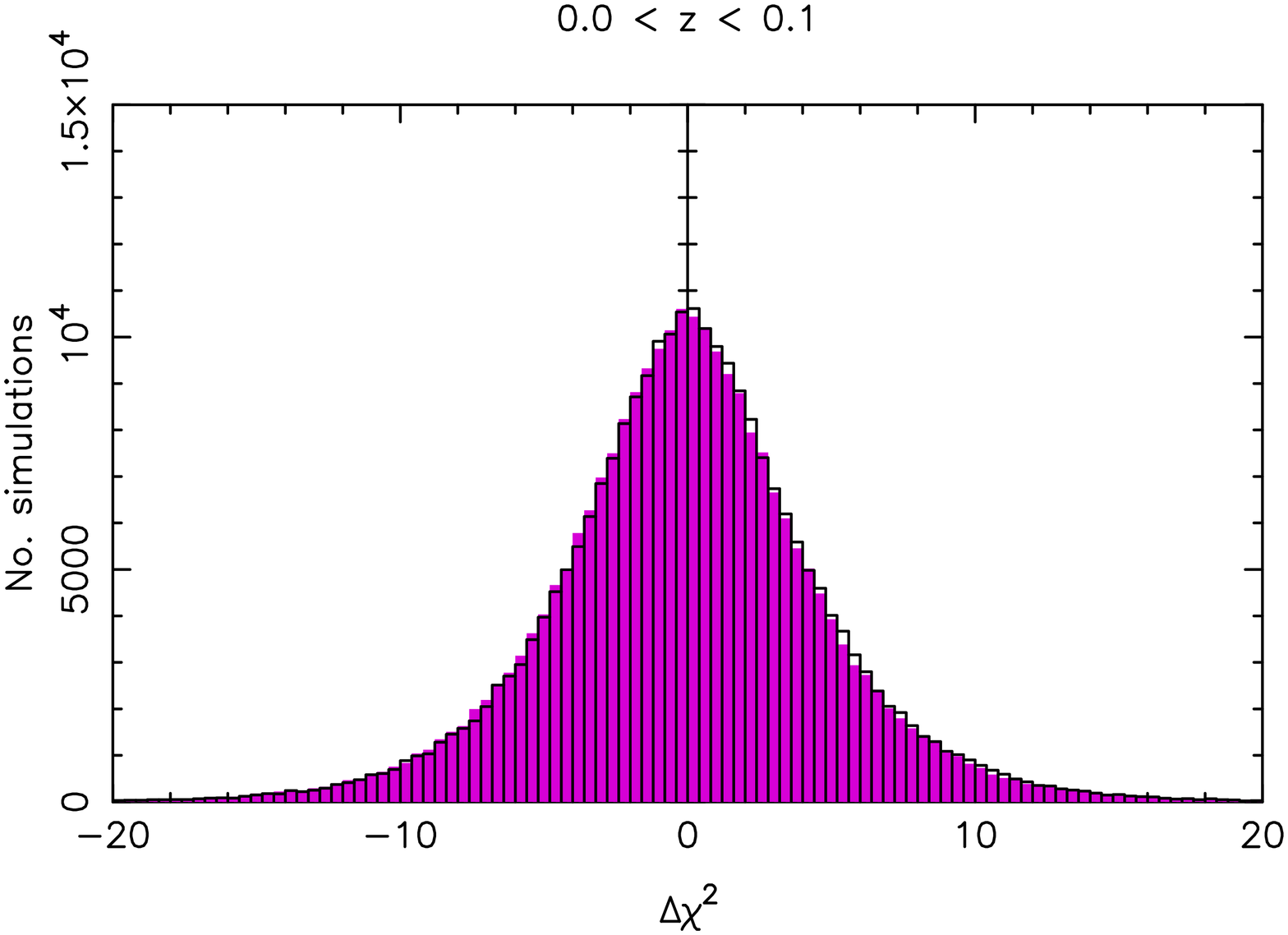,width=6.0cm,height=8.0cm}
}
\end{minipage}
\hspace{0.5cm}
\begin{minipage}[r]{7.0cm}
\centering
\rotatebox{270}{
\epsfig{file=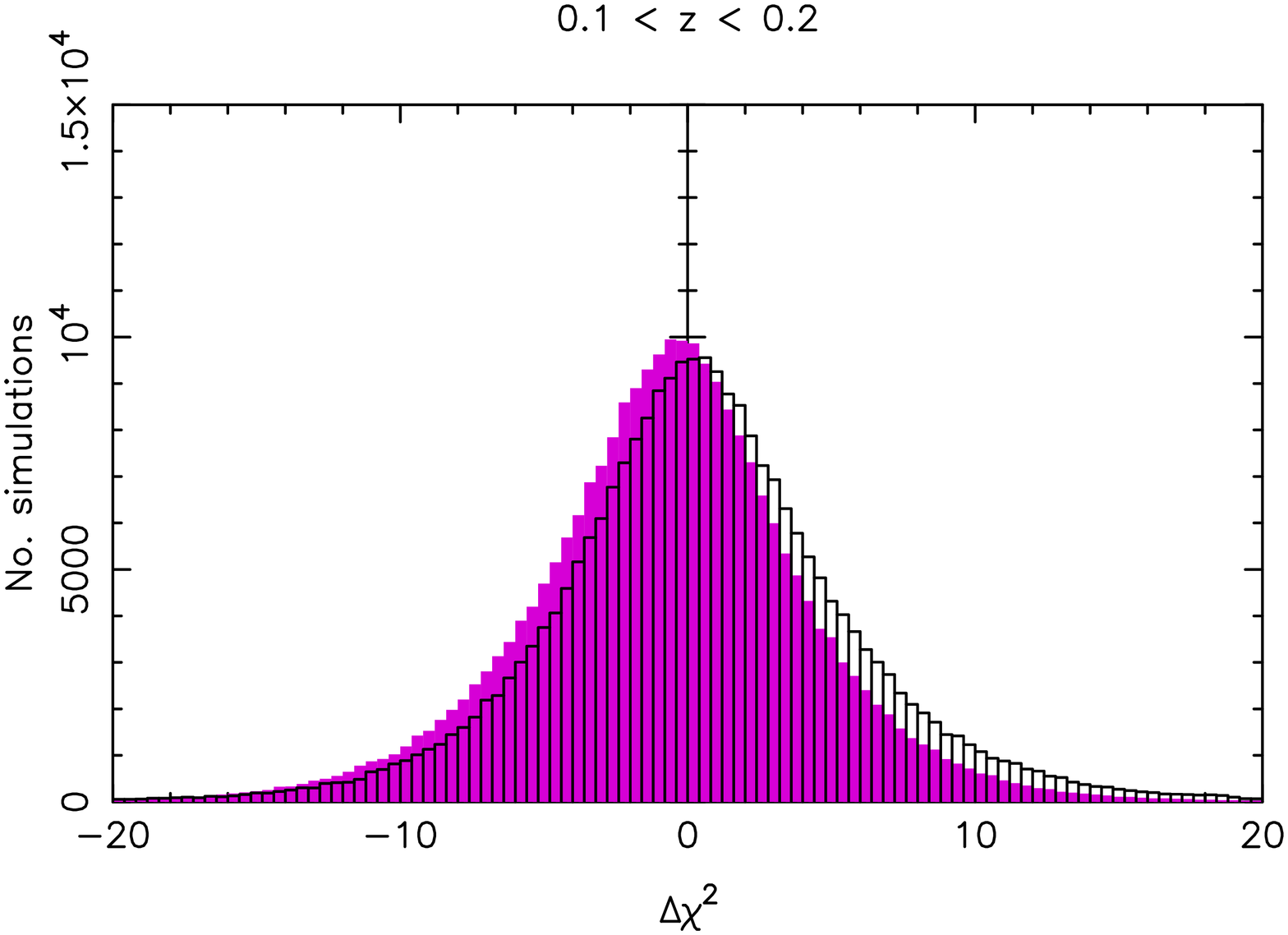,width=6.0cm,height=8.0cm}
}
\end{minipage}

\begin{minipage}[l]{7.0cm}
\centering
\rotatebox{270}{
\epsfig{file=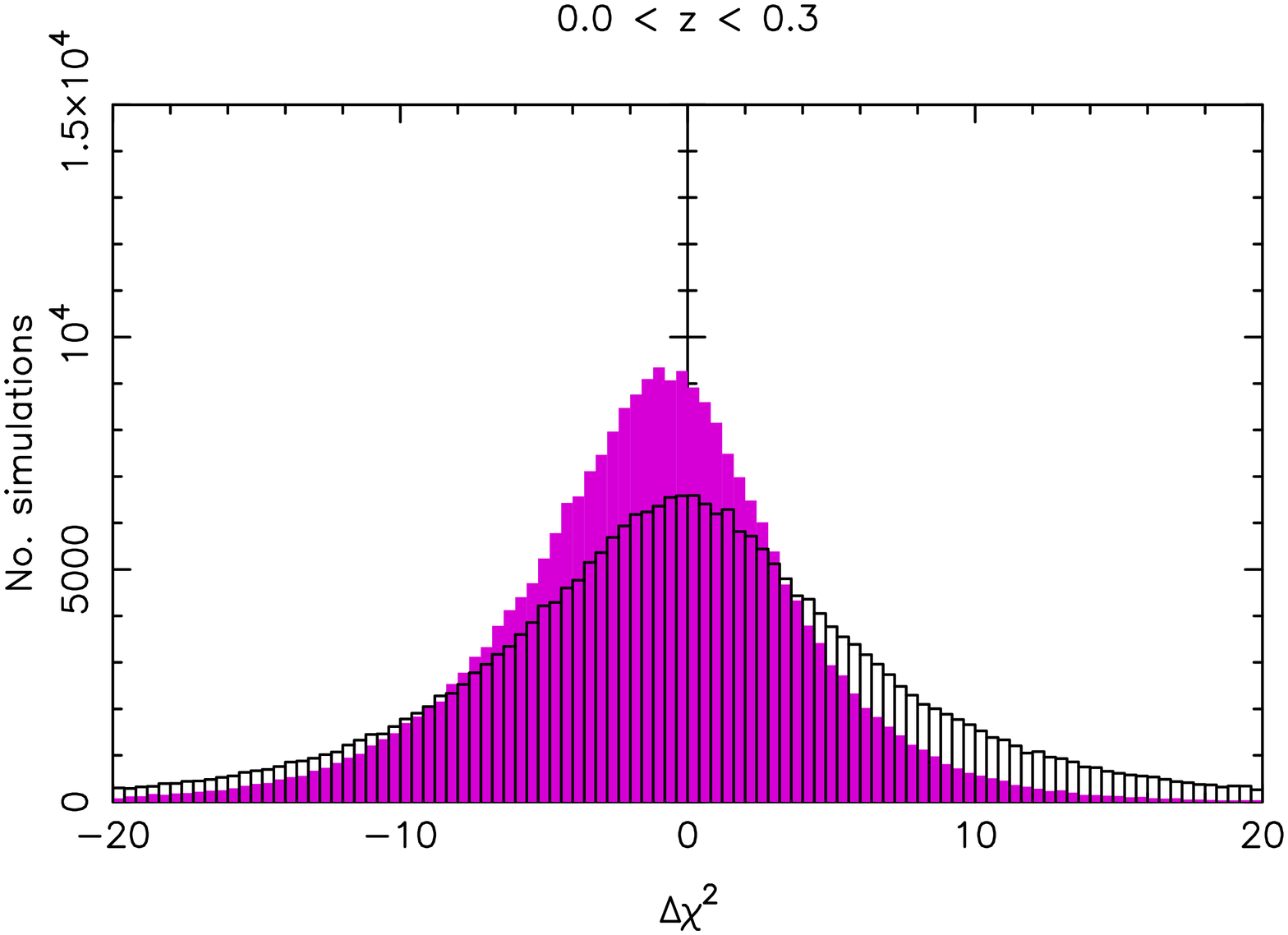,width=6.0cm,height=8.0cm}
}
\end{minipage}
\hspace{0.5cm}
\begin{minipage}[r]{7.0cm}
\centering
\rotatebox{270}{
\epsfig{file=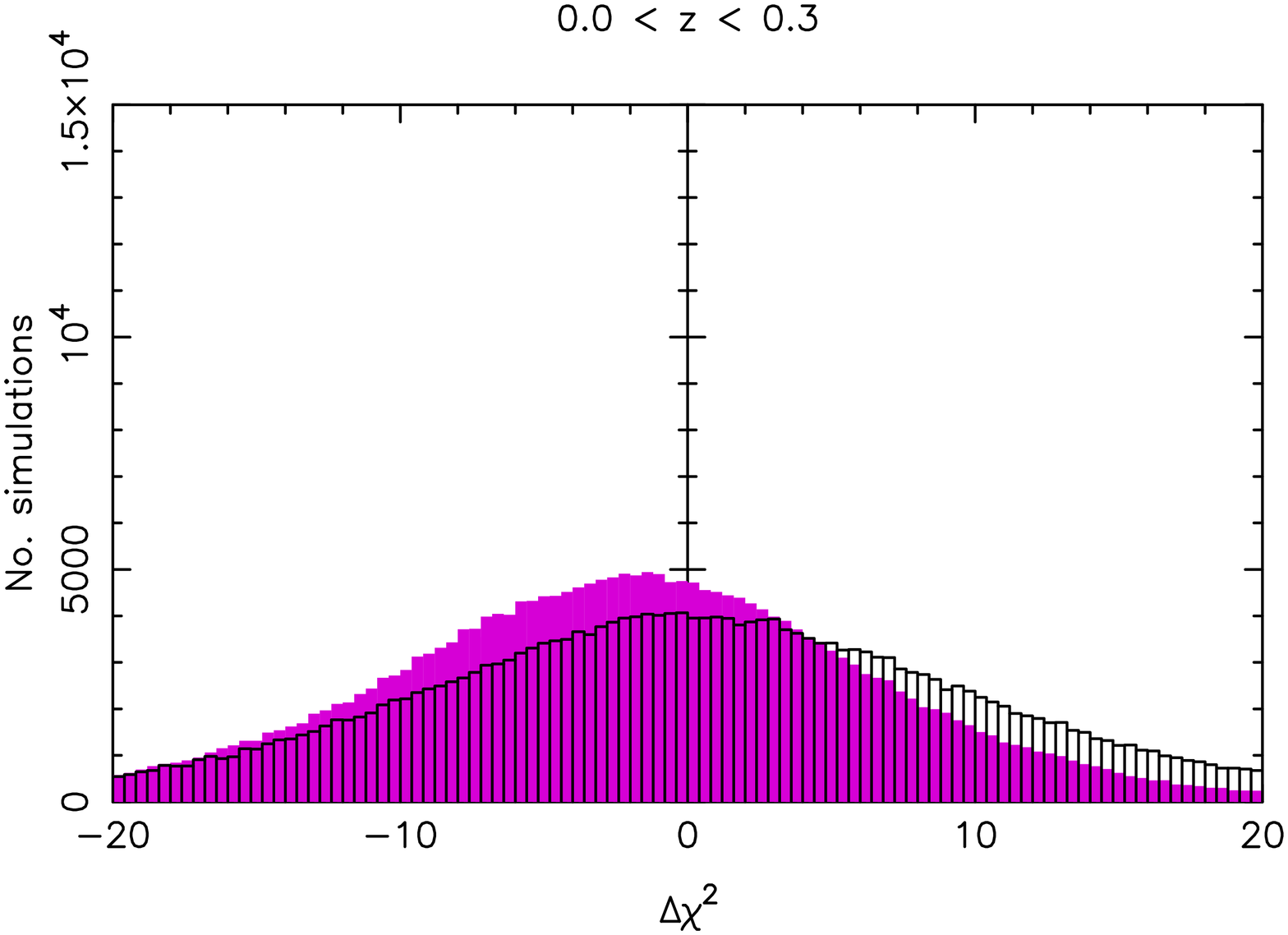,width=6.0cm,height=8.0cm}
}
\end{minipage}
\caption{The filled histograms show the values of $\Delta\chi^{2}$
  calculated from simulation for the null hypothesis (no ISW effect);
  the outline histograms are for the alternative hypothesis
  ($\Lambda$CDM). Panels are $0.0<z<0.1$ (top left), $0.1<z<0.2$ (top
  right), $0.2<z<0.3$ (bottom left) and the full data set $0.0<z<0.3$
  (bottom right). A positive $\Delta\chi^{2}$ indicates that the
  $\Lambda$CDM signal is preferred. Thus a powerful experiment would
  generate negative values of $\Delta\chi^{2}$ if the null hypothesis
  is simulated, and positive values if $\Lambda$CDM is simulated. In
  practice, we see little offset between the $\Delta\chi^{2}$
  distributions in all cases. As we move to higher redshift slices,
  the peaks of the $\Delta\chi^{2}$ distributions begin to separate
  slightly, although any offset remains small. The broader histograms
  in the $0.2<z<0.3$ slice are a result of shot noise in the galaxy
  density field.\label{fig:delta_chi_data}}
\end{figure*}

The set of simulated galaxy and CMB+ISW maps allows us to calculate
the distribution of the $\Delta\chi^{2}$ values we should expect if
the alternative hypothesis were true:
\[
P(\Delta\chi^{2} = x \mid H_{1} \textrm{ true}).
\]
We can thus measure the probability of making a Type II error
(accepting the null hypothesis when the alternative is true) for a
given threshold. Fig. \ref{fig:delta_chi_data} shows the two
$\Delta\chi^{2}$ histograms computed from simulations of the CMB and
2MASS galaxy data for the different redshift slices used in the
analysis.

A powerful statistical test would show a clear offset between the
$\Delta\chi^{2}$ histograms in Fig. \ref{fig:delta_chi_data}. In
reality, we see very little offset between the $\Delta\chi^{2}$
distributions for this experiment. The final panel of
Fig. \ref{fig:delta_chi_data} reveals that if the $\Lambda$CDM
hypothesis is true, a very large value of $\Delta\chi^{2}$ would be
needed to rule out the null hypothesis with any confidence. There is
an $8\%$ chance of obtaining a $\Delta\chi^{2}$ value greater than
$15$ if the $\Lambda$CDM hypothesis is true, and a $5\%$ chance of
making a Type I error with this threshold.

If we were to adopt Jeffreys' criterion \citep{Jeffreys_1948} that
$\Delta\chi^{2}>5$ constitutes strong evidence for rejecting the null
hypothesis, then such a threshold carries a $\sim 23\%$ chance of
making a Type I error.


To make a decisive detection of the ISW effect we need
$\Delta\chi^{2}$ distributions with a clearer offset and less
overlap. The poor 2MASS results prompt us to ask how good the data
would have to be to make a significant ISW detection. As the CMB data
on these scales are already signal-dominated, any improvement must
come from the galaxy data.

The \emph{power} of a statistical test is defined as
\[
1-P(s_{*})
\]
where $P(s_{*})$ is the probability of making a Type II error with a
threshold $s_{*}$. A statistical test can be made arbitrarily powerful
by altering the threshold, but a gain in power is offset by an
increase in the probability of making a Type I error. We therefore
define the `Optimal Power' of a test as the power of that test with
threshold chosen such that the probabilities of making Type I and Type
II errors are equal. A larger Optimal Power indicates a `better'
statistical test in the sense that it is less likely to return the
wrong conclusion from the data. The Optimal Power of this experiment
is $0.55$.

\subsection{Same survey, fainter magnitude limit}

\label{sec:moregals}

Consider a survey with the same parameters as above but more galaxies
at higher redshift ($0.1<z<0.3$), equivalent to having a fainter
magnitude limit for the survey. Following the same procedure used in
Section \ref{sec:covmat}, we simulate lognormal galaxy maps for this
survey with bias and number density equal to the measured values for
$z<0.1$.

The $\Lambda$CDM $\Delta\chi^{2}$ distribution in the third redshift
slice (histograms not shown here) is narrower than that for the 2MASS
experiment, although the `null' distribution is largely unchanged,
reflecting a reduction in shot noise which improves the ISW signal
estimation. Looking at the analysis of the entire dataset together,
the peak of the $\Lambda$CDM $\Delta\chi^{2}$ distribution is shifted
to a higher value and the distribution is narrower, leading to an
increase in the offset between the peaks of the
distributions. However, the Optimal Power for this experiment is
$0.60$, showing little improvement.



\subsection{Deeper survey (more redshift slices)}

\label{sec:deeper}

We also investigate hypothetical deeper surveys, with characteristic
redshift $z_{\mathrm{m}}$. The redshift distribution is computed using
\[
n(z) = n_{\mathrm{com}}(z)\exp\{-(z/z_{m})^{1.6}\},
\]
where $n_{\textrm{\scriptsize com}}$ is constant and calculated from
2MASS galaxies with $z<0.1$. We also consider varying
$z_{\textrm{\scriptsize max}}$. Generally, we shall show results
for a maximum redshift up to 0.7.

We consider shells of width $\Delta z = 0.1$ and take the true
redshift distribution of galaxies in each slice as a Gaussian of width
$\sigma = 0.03(1+z)$ centred at the midpoint of the slice for $z>0.3$;
the true redshift distributions calculated for the 2MASS data are used
for $z<0.3$. The galaxy bias in each slice is estimated using the
simple assumption that bias is determined solely by the ratio of
observed to expected numbers of galaxies in that slice, where the
expected number of galaxies is found using $n_{\textrm{\scriptsize
com}}$. This means that bias is not intrinsically redshift dependent
-- the reason we see higher bias at higher redshift is because only
the brightest, most massive galaxies are observed there. This seems a
reasonable model for the redshifts we consider given the results of
\citet{Magliocchetti}, who see little evolution in bias over this
range. In keeping with these assumptions, we use the bias in the three
2MASS redshift slices, together with the fraction of galaxies observed
here, to deduce a relationship between these quantities. Our
best-fitting relationship is
\[
b = 1.2f_{\textrm{\scriptsize gals}}^{-0.13}, \label{eq:biaspred}
\]
where $f_{\mathrm{gals}}$ is the fraction of
galaxies observed.

\begin{figure*}
\centering
\begin{minipage}[l]{7.0cm}
\centering
\rotatebox{270}{
\epsfig{file=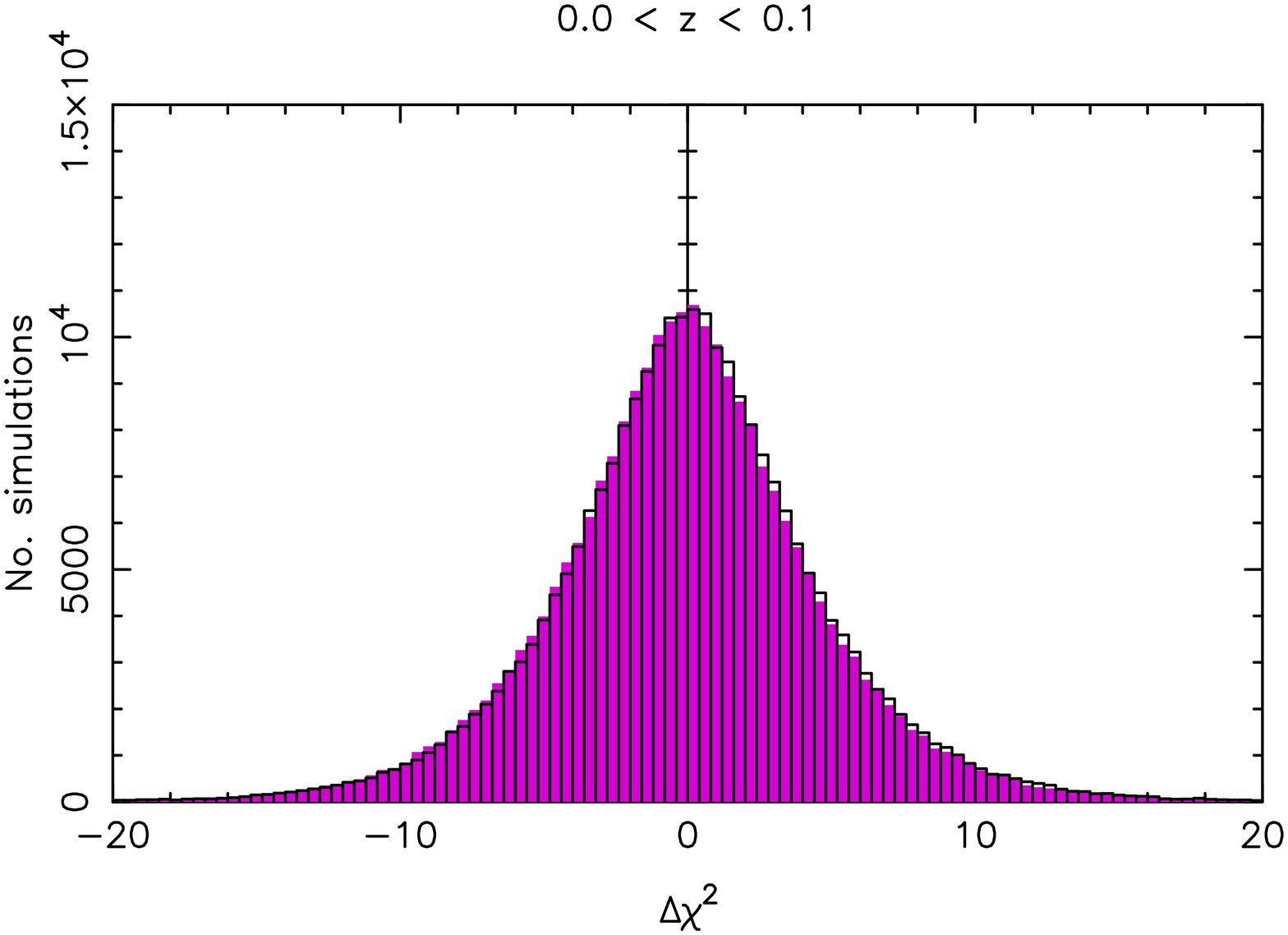,width=5.5cm,height=7.5cm}
}
\end{minipage}
\hspace{0.5cm}
\begin{minipage}[r]{7.0cm}
\centering
\rotatebox{270}{
\epsfig{file=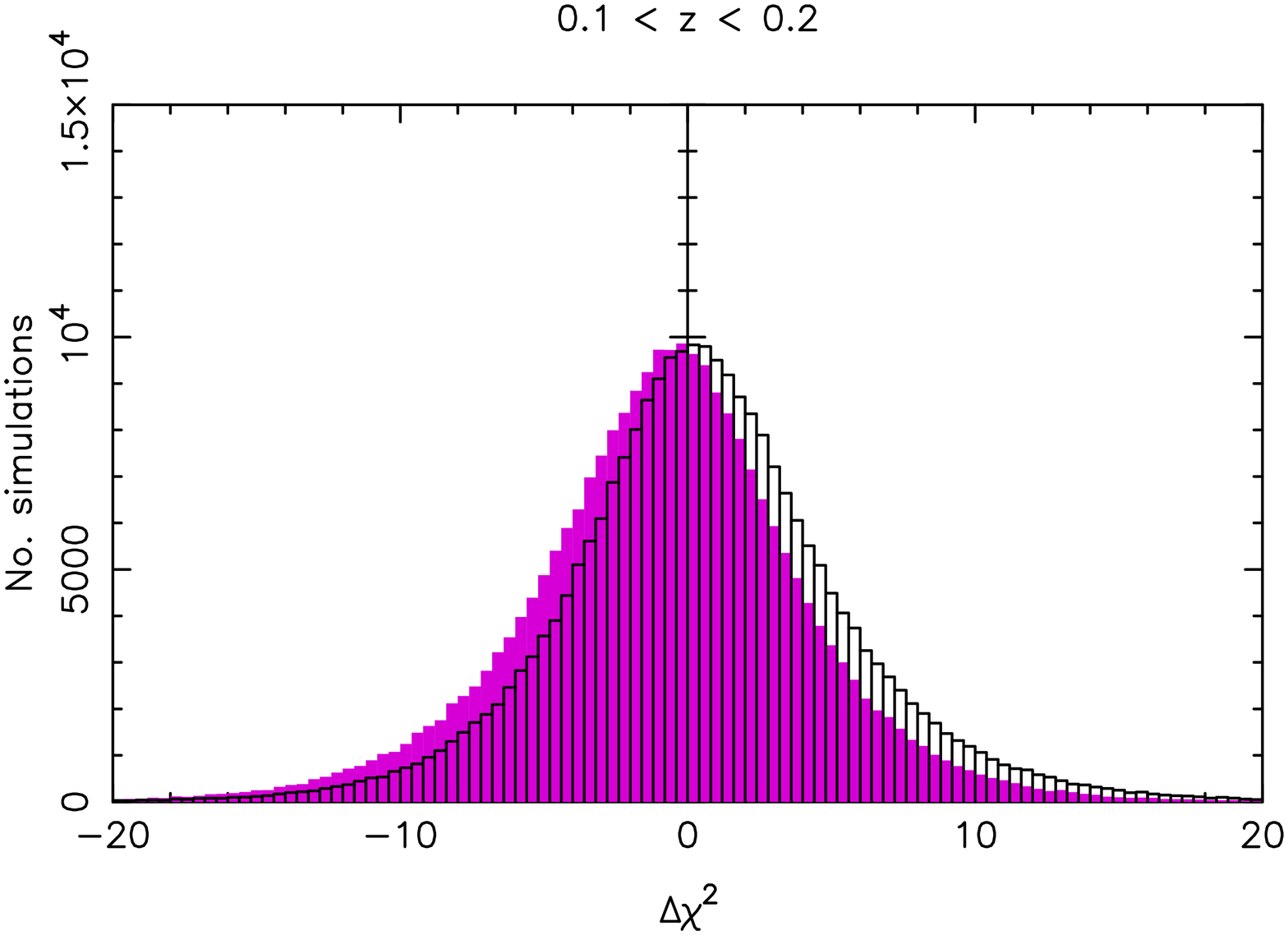,width=5.5cm,height=7.5cm}
}
\end{minipage}

\vspace{-5pt}

\begin{minipage}[l]{7.0cm}
\centering
\rotatebox{270}{
\epsfig{file=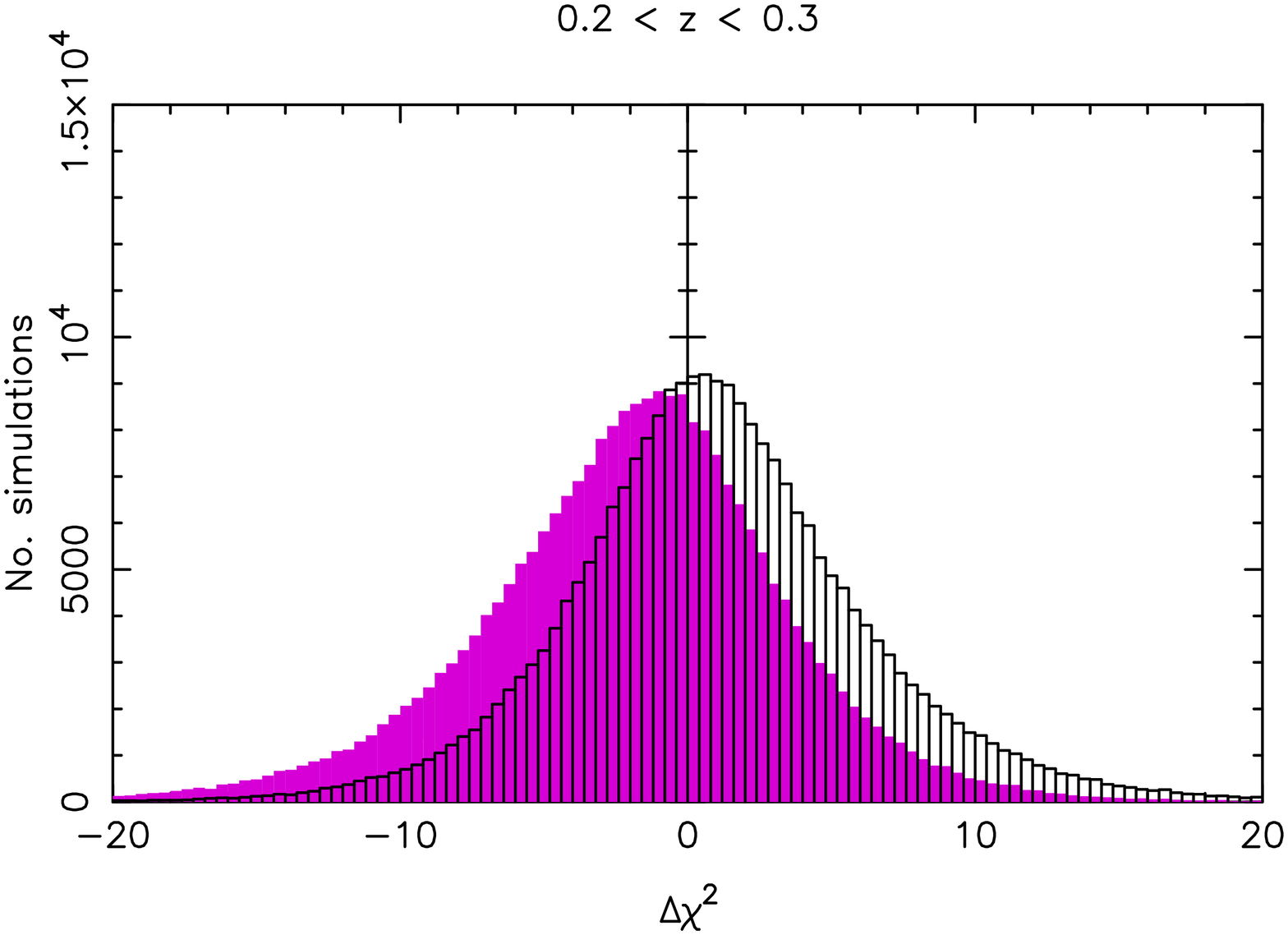,width=5.5cm,height=7.5cm}
}
\end{minipage}
\hspace{0.5cm}
\begin{minipage}[r]{7.0cm}
\centering
\rotatebox{270}{
\epsfig{file=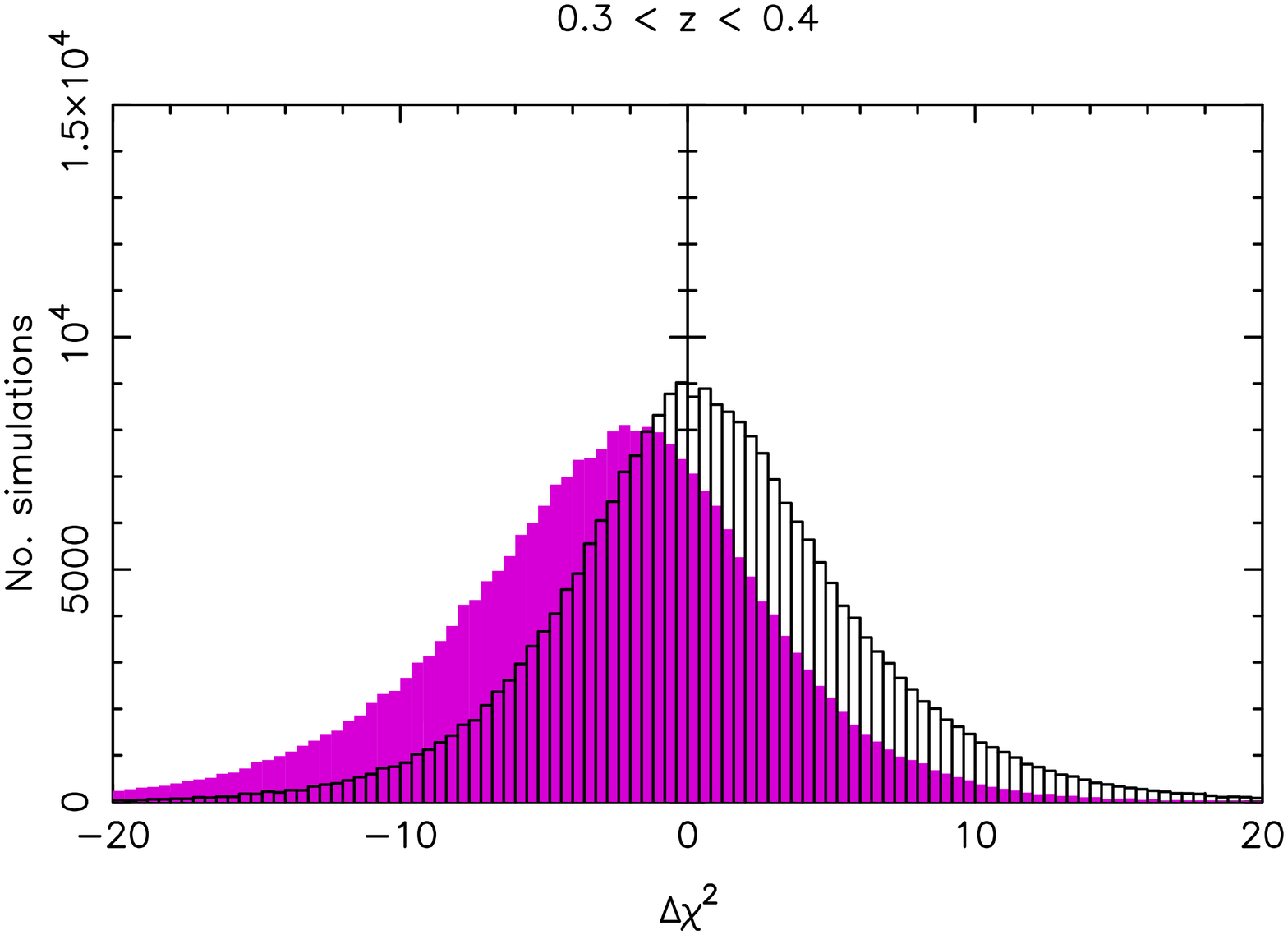,width=5.5cm,height=7.5cm}
}
\end{minipage}

\vspace{-5pt}

\begin{minipage}[l]{7.0cm}
\centering
\rotatebox{270}{
\epsfig{file=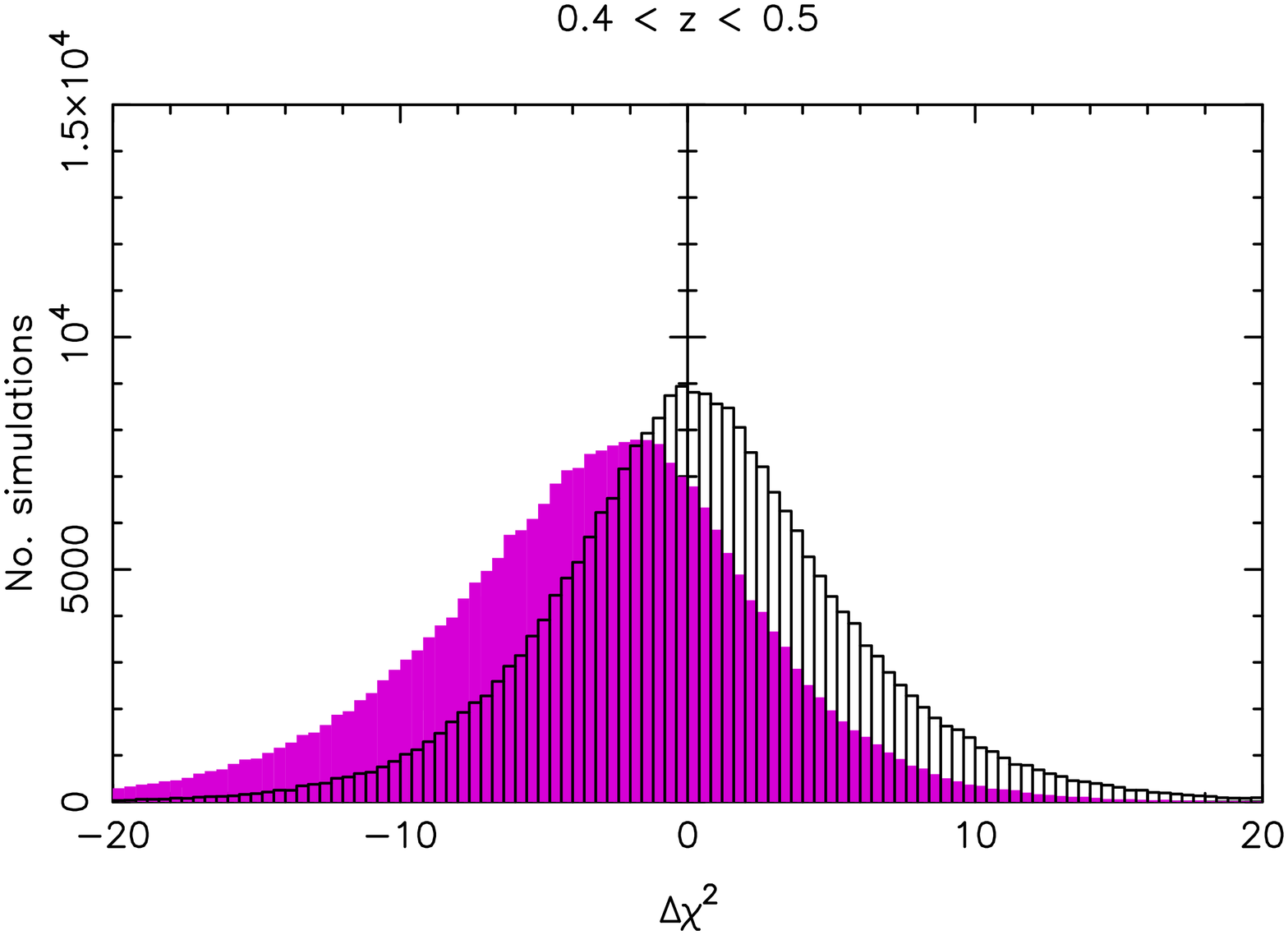,width=5.5cm,height=7.5cm}
}
\end{minipage}
\hspace{0.5cm}
\begin{minipage}[r]{7.0cm}
\centering
\rotatebox{270}{
\epsfig{file=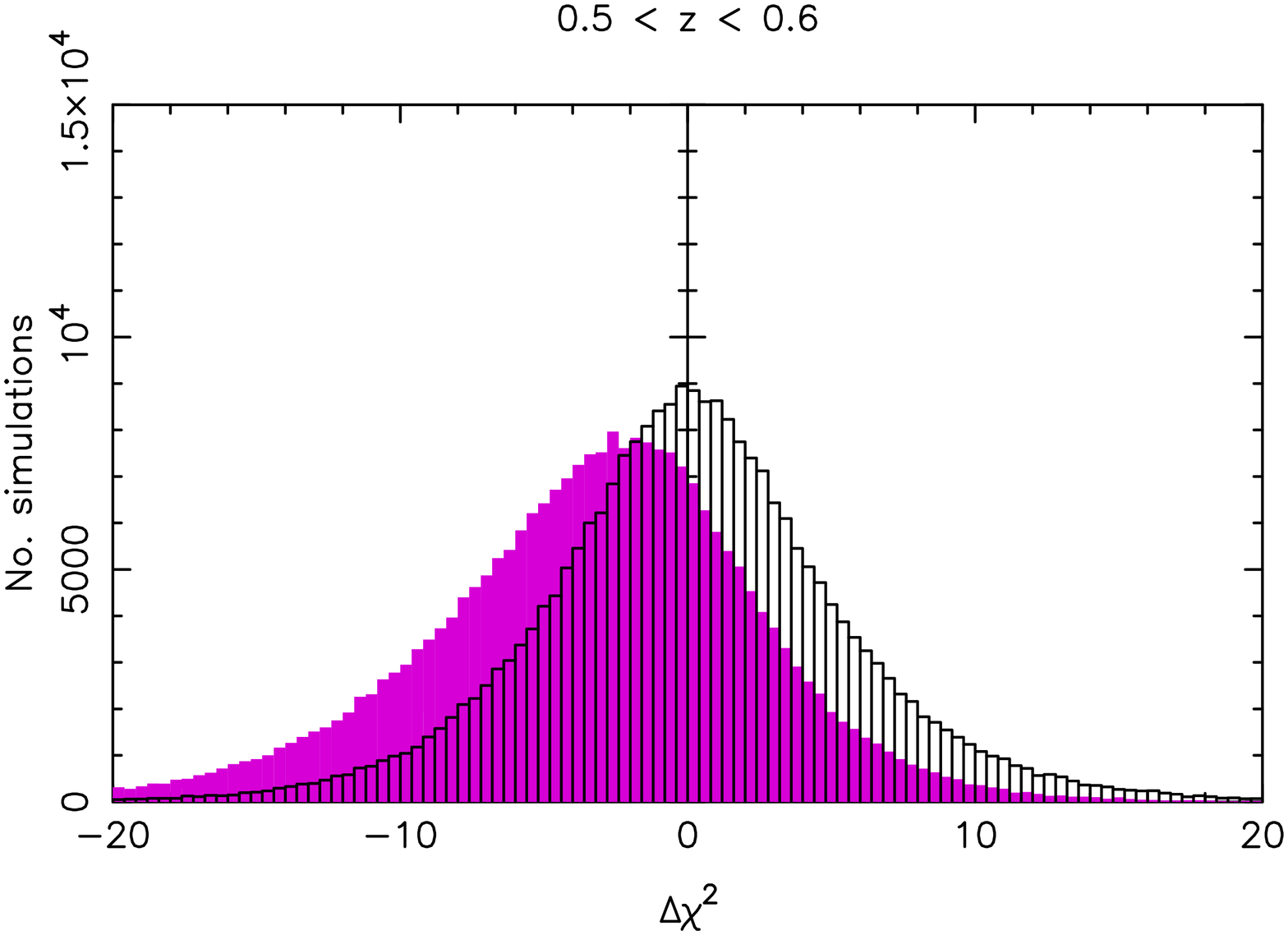,width=5.5cm,height=7.5cm}
}
\end{minipage}

\vspace{-5pt}

\begin{minipage}[l]{7.0cm}
\centering
\rotatebox{270}{
\epsfig{file=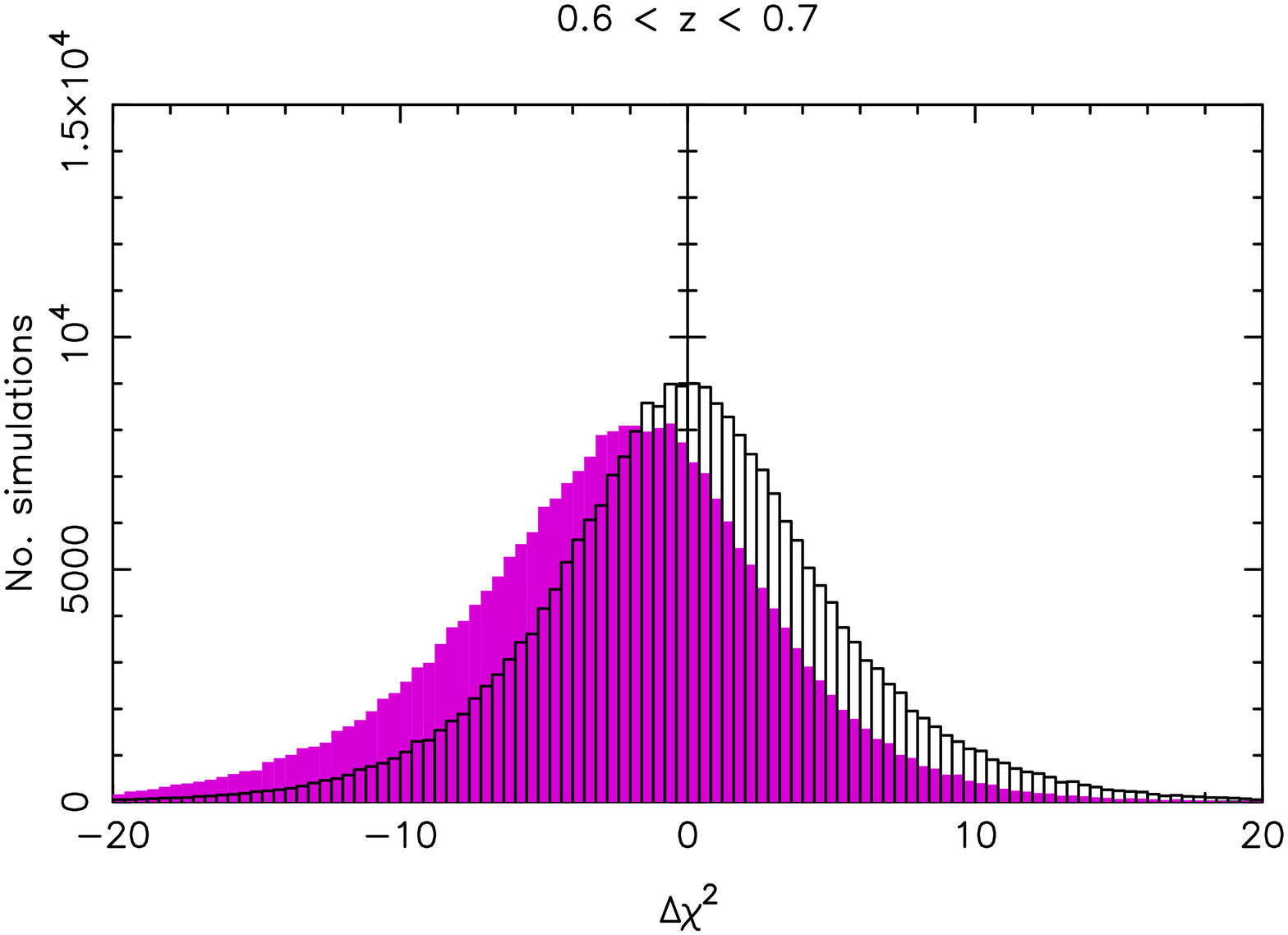,width=5.5cm,height=7.5cm}
}
\end{minipage}
\hspace{0.5cm}
\begin{minipage}[r]{7.0cm}
\centering
\rotatebox{270}{
\epsfig{file=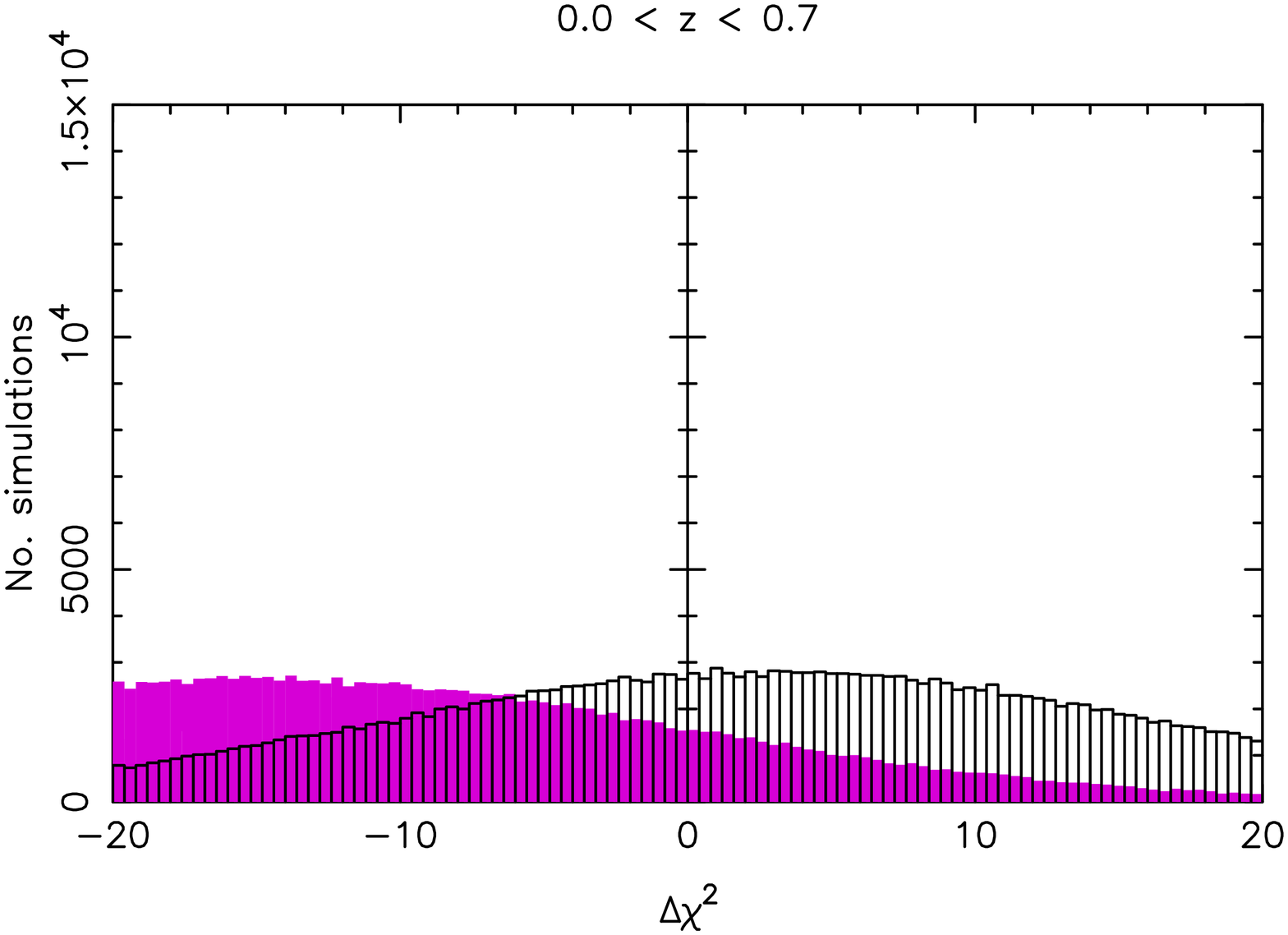,width=5.5cm,height=7.5cm}
}
\end{minipage}
\caption{A simulated ISW detection experiment with characteristic
  depth $z_{\mathrm{m}}=0.3$ and $z_{\mathrm{max}}=0.7$. The filled
  histograms show the values of $\Delta\chi^{2}$ calculated for the
  null hypothesis; the outline histograms are for the values of
  $\Delta\chi^{2}$ for the $\Lambda$CDM hypothesis. Comparison of the
  first three redshift slices with the previous 2MASS analysis shows
  improved detection prospects for these slices and
  overall. \label{fig:delta_chi_z0.3}}
\end{figure*}

Fig. \ref{fig:delta_chi_z0.3} shows the $\Delta\chi^{2}$ histograms
for this experiment with $z_{\mathrm{m}}=0.3$. The $\Delta\chi^{2}$
distributions for redshift slices with $z>0.3$ all show an offset and
the null hypothesis distributions for these slices are slightly skewed
towards negative values. The analysis of the full dataset for
$0.0<z<0.7$ shows a clear offset between the peaks of the two
distributions and considerable improvement in detection prospects.

\begin{figure}
\centering
\epsfig{file=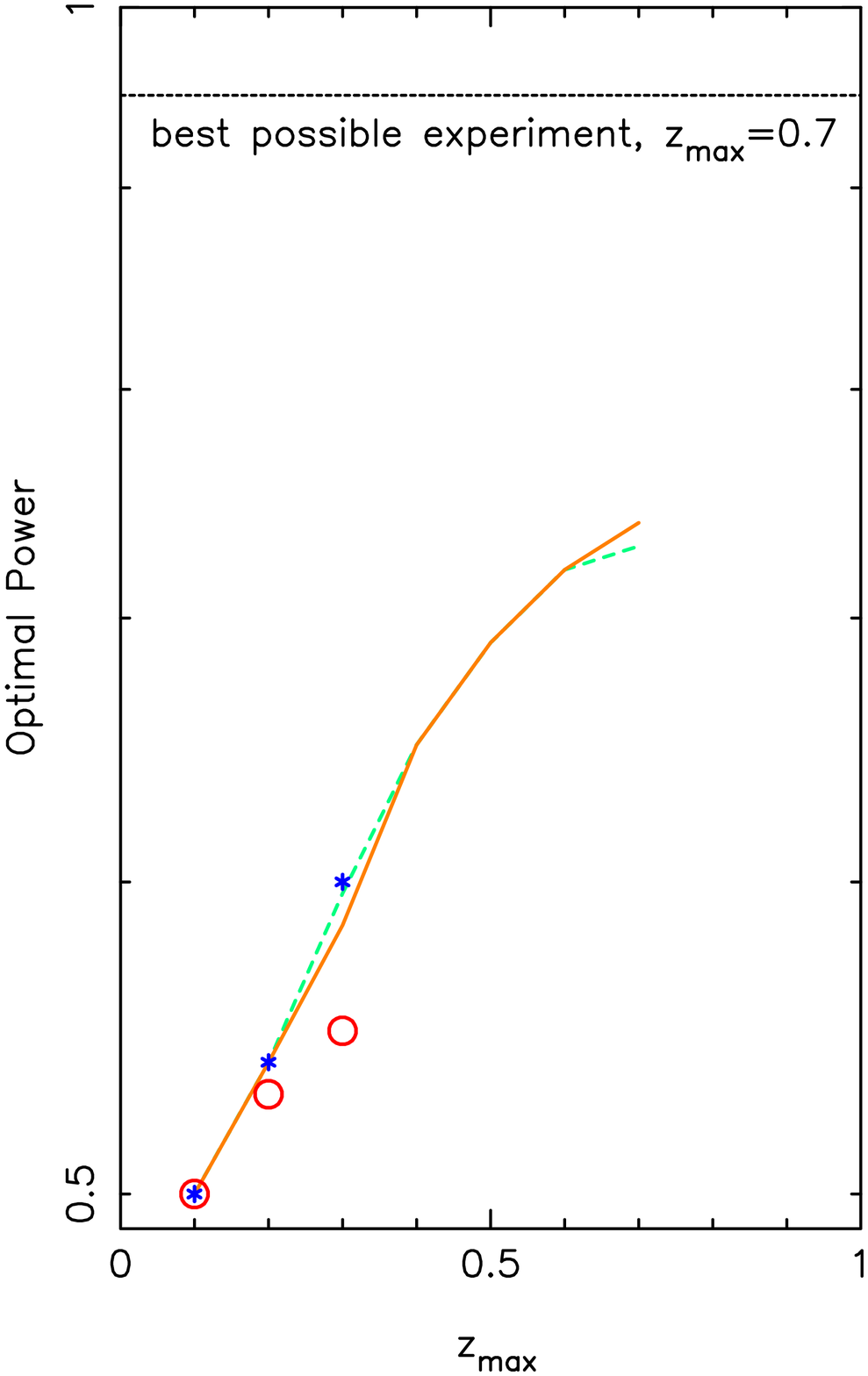,width= 8.0cm,height=9.0cm}
\caption{The build up of Optimal Power as a function of
$z_{\textrm{\scriptsize max}}$ for the 2MASS experiment (red circles),
more galaxies (blue stars), $z_{\mathrm{m}}=0.3$ (green dashed line)
and $z_{\mathrm{m}}=0.75$ (orange solid
line). The top horizontal line shows the idealized case where
the ISW signal (for $z_{\textrm{\scriptsize max}}=0.7$ is known
exactly, rather than estimated from galaxy data.
\label{fig:power_buildup}}
\end{figure}

Values of $\Delta\chi^{2} \gtrsim 10$ would be expected to occur (if
the alternative hypothesis were true) around $30\%$ of the time and
would lead to a fairly conclusive rejection of the null hypothesis
($5\%$ chance of a Type I error). The Jeffreys criterion of
$\Delta\chi^{2}>5$ giving a `strong' detection would have a $9\%$
chance of a Type I error and would be satisfied by the data $43\%$ of
the time under the $\Lambda$CDM hypothesis. The Optimal power of this
experiment is $0.73$; Fig. \ref{fig:power_buildup} shows how this
increases with depth.

The results for an experiment with $z_{\mathrm{m}}=0.7$ are
practically indistinguishable from those for the $z_{\mathrm{m}}=0.3$
survey and are therefore not shown in full. The only difference is a
slight increase in offset in the $\Delta\chi^{2}$ distributions for
$0.5<z<0.6$ and $0.6<z<0.7$. This is caused by a reduction in shot
noise in these slices leading to a better estimation of the ISW
signal. The Optimal Power of this survey is $0.74$, its build up as
successive redshift slices are added to the analysis is again shown in
Fig. \ref{fig:power_buildup}.

The fact that the power depends mainly on $z_{\mathrm{max}}$ and
hardly on $z_{\mathrm{m}}$ may at first seem surprising. But as long
as there are enough galaxies observed in a redshift slice that shot
noise is small, and the bias in the slice is known, the number of
galaxies and the particular value of $b$ are unimportant.

\subsection{Results for the `ideal' case}

\begin{figure}
\centering
\rotatebox{270}{
\epsfig{file=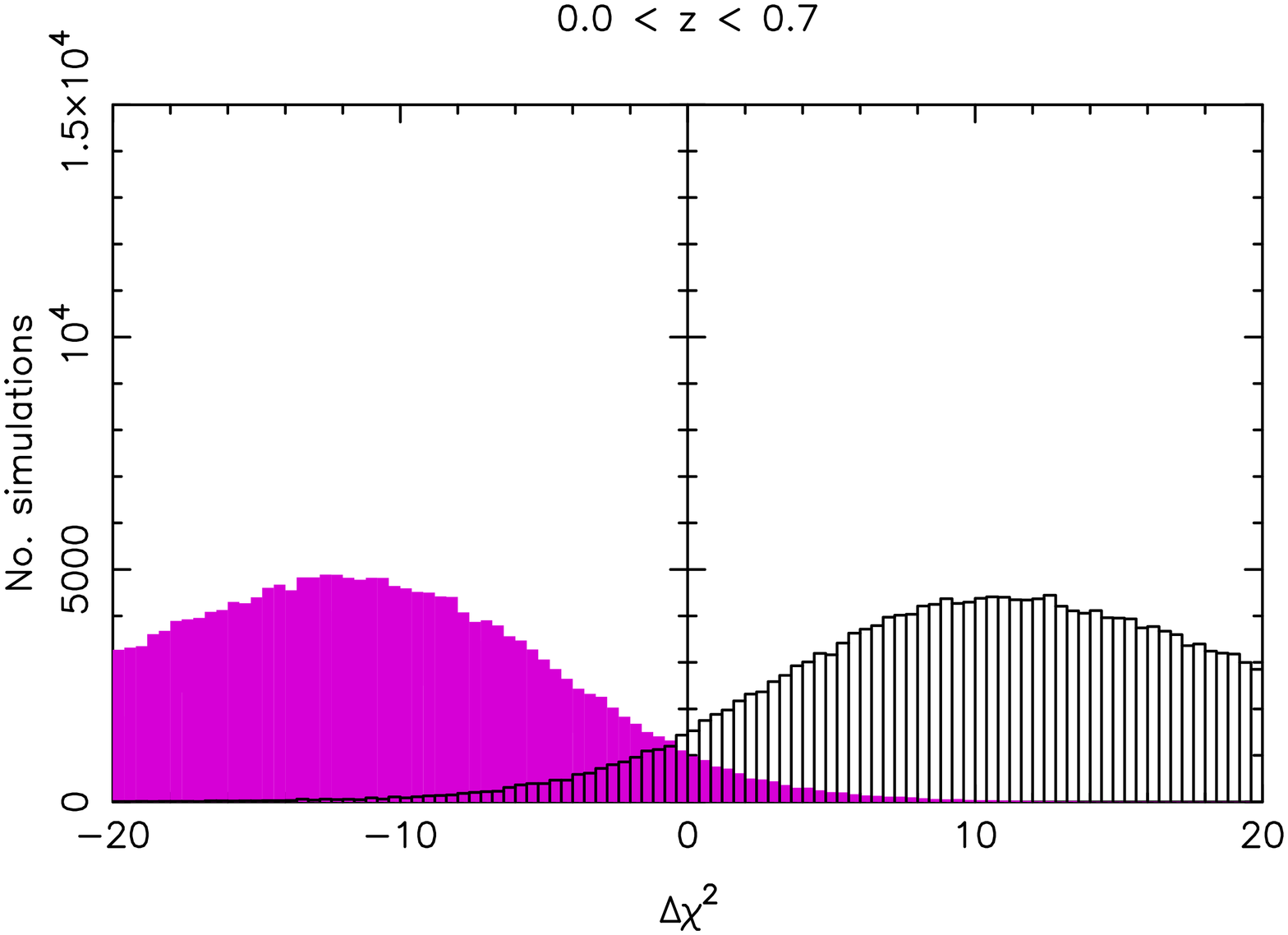,height=\linewidth,width=6.0cm}
}
\caption{The filled histogram shows the $\Delta\chi^{2}$ values for
  the null hypothesis in the `ideal' case where we could directly
  measure the ISW temperature fluctuations to $z_{\mathrm{max}}=0.7$;
  the outline histogram is for the $\Lambda$CDM hypothesis. We see a
  much diminished overlap between the distributions in this ideal
  case, although cosmic variance means that there is not complete
  separation. \label{fig:allsky_ideal_zm0.7}}

\rotatebox{270}{
\epsfig{file=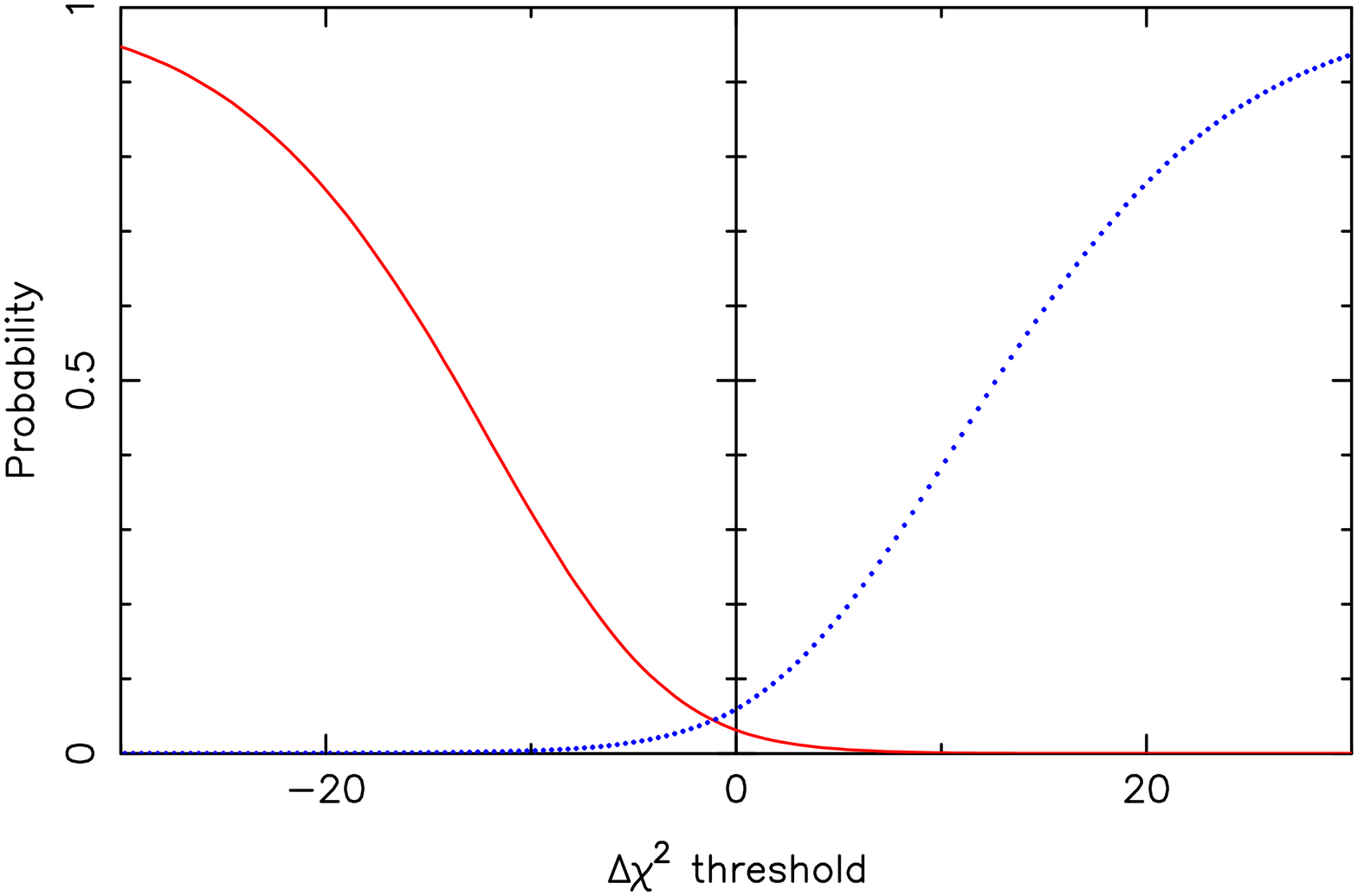,width= 6.0cm,height=\linewidth}
}
\caption{The probability of making a Type I error (solid line) and a
  Type II error (dotted line) as a function of threshold in
  $\Delta\chi^{2}$ for the ideal experiment with
  $z_{\mathrm{max}}=0.7$. The Optimal Power of this experiment (0.95)
  is $1-p$ where $p$ is the value of the probability where the above
  lines cross. We note that under the $\Lambda$CDM hypothesis, there
  is a $\sim 15\%$ chance of not detecting the ISW effect with a
  threshold for detection of $\Delta\chi^{2} \ge
  5$. \label{fig:type1_2_ideal}}
\end{figure}

The best conceivable ISW experiment would measure directly the
temperature fluctuations due to the ISW effect (e.g. from perfect
knowledge of the dark matter density field) across the entire sky, and
cross-correlate such maps with the all-sky CMB. We consider this
hypothetical limiting case in terms of our $\Delta\chi^{2}$ analysis
using realizations of the ISW angular power spectrum as our `galaxy'
maps. Even in this `best possible' case, we do not expect to be able
to reject the null hypothesis with absolute certainty due to cosmic
variance on the scales of the ISW effect. The $\Delta\chi^{2}$
histograms we find for the ISW effect with $z_{\mathrm{max}}=0.7$ are
shown in Fig. \ref{fig:allsky_ideal_zm0.7} and the probabilities of
making Type I and II errors as a function of threshold in
Fig. \ref{fig:type1_2_ideal}. We notice that the separation of the
distributions is fairly distinct, although a non-negligible overlap remains;
the Optimal Power for this experiment is 0.95. 

Interestingly, if $\Delta\chi^{2} \ge 5$ constitutes `strong' evidence
for the ISW effect in a $\Lambda$CDM universe, then we would expect to
fail this criterion $\sim 15\%$ of the time i.e. only $85\%$ of
$\Lambda$CDM universes have an ISW effect that is detectable at this
level. In the remaining $\sim 15\%$, the intrinsic CMB signal serves
to effectively `hide' the ISW signal. Bear in mind that this
experiment assumes all-sky measurement of the ISW temperature
fluctuations to $z_{\mathrm{max}}=0.7$, and that
(necessary) masking of the sky will, as shown earlier, degrade
the power of the experiment with any realistic galaxy data, so
that probably 1 in 4 observers would be incapable of detecting
the ISW effect given data to $z_{\mathrm{max}}=0.7$.

This redshift limit of 0.7 includes about 50\% of the ISW
power compared to $z_{\mathrm{max}}=\infty$. One can therefore do
better with a larger redshift limit (e.g. $z_{\mathrm{max}}=1.3$
would capture 85\% of the total power). But even in the limiting case
of perfect knowledge of the total ISW signal back
to last scattering over the whole sky, we would fail to
find `strong' $\Delta\chi^{2} \ge 5$ evidence for the ISW
effect in 2\% of $\Lambda$CDM universes. With the combined practical
effects of realistic redshift coverage, galaxy sampling, and sky
masking for the best foreseeable future surveys, this figure
would rise to the order of 10\%.
Our results thus suggest that it is not
unlikely that we could inhabit a universe where the ISW effect is
present but undetectable.

\section{Conclusions}

\label{sec:conclusions}

We have analysed photometric redshift data from 2MASS to
$z_{\textrm{\scriptsize max}}=0.3$ and WMAP CMB data to look for a
cross-correlation signal indicative of an ISW effect in a $\Lambda$CDM
universe. This builds on the work of R07 who perform a similar
analysis without redshift information. Our results are equally
inconclusive: the data do not rule out a null hypothesis of no ISW
effect; a $\Lambda$CDM ISW signal is preferred, but with a likelihood ratio of
only $1.5:1$. We use error bars computed from simulations of both CMB skies
\emph{and} lognormal galaxy density fields and impose a slightly
stricter magnitude cut than R07 on the 2MASS data to ensure
uniformity. The smaller sample size and larger error bars that we
employ can explain the fact that our $\chi^{2}$ values are slightly
more unusually low than those of R07, but the generally
low $\chi^{2}$ nature of this dataset remains.

In the light of our inconclusive results, we have considered the
expected power of several ISW detection experiments. As it turns out,
2MASS is expected to be poor at discriminating between the two
hypotheses considered. Simulations of hypothetical deeper photometric
redshift surveys reveal detection power to mainly depend upon the
value of $z_{\mathrm{max}}$ used for the analysis rather than the size
of the galaxy sample. As long as sufficient galaxies are observed in each
redshift slice to keep shot noise from dominating, the precise number of
galaxies is largely irrelevant.

The limiting case for such detection experiments is that of the
`ideal' ISW experiment where one can measure the ISW temperature
fluctuations precisely out to a given $z_{\mathrm{max}}$. Even in this
over-idealized case, we do not expect
perfect detection prospects due to cosmic variance on these
scales: a certain fraction of intrinsic CMB
skies have large-scale temperature fluctuations that `hide' the
ISW effect. Simulations indicate that, in this ideal case, a Jeffreys' criterion of
$\Delta\chi^{2} = \chi^{2}_{\mathrm{null}}-\chi^{2}_{\mathrm{fid}} \ge
5$ for a `strong' detection would fail to be satisfied $\sim 10\%$ of the time, for
the best conceivable practical datasets.

Given that the ISW effect seems doomed to be a low S/N phenomenon, is
there a conflict with existing studies that claim to have detected the
effect at high levels of significance? Not necessarily; our results
concern how frequently one will fail to find compelling evidence for
the ISW effect, and do not say how strong the evidence will be if we
are lucky enough to live in a universe where this detection is
possible. Nevertheless, there is reason to doubt how much further the
ISW effect can be pushed as a probe of precision cosmology. In terms
of sky coverage and redshift range, most of the ISW signal remains to
be mapped. Even if we have been lucky with the regions of space
studied to date, there is no guarantee that this will apply to the
total signal. Future studies of the ISW effect may fail to increase
the significance of current detections, and could even reduce them;
the recent non-detection by \citet{Sawangwit_ISW10} using a
high-redshift LRG sample may be an example of this phenomenon.

\section*{Acknowledgements}

CLF was supported by a PPARC postgraduate studentship. We thank Kate
Land for much helpful correspondence. This research has made use of
optical data obtained from the SuperCOSMOS Science Archive, prepared
and hosted by the IfA's Wide Field Astronomy Unit, consisting of
scanned survey plates from the UK Schmidt Telescope and The Palomar
Observatory Sky Survey (POSS-II). This publication also makes use of
data products from the Two Micron All Sky Survey, which is a joint
project of the University of Massachusetts and the Infrared Processing
and Analysis Center/California Institute of Technology, funded by the
National Aeronautics and Space Administration and the National Science
Foundation.  \phantom{\citet{Gorski_HEALPix},  \citet{Cooray_and_Sheth2002}}

\setlength{\bibhang}{2.0em}
\setlength\labelwidth{0.0em}
\bibliography{biblio}

\begin{thebibliography}{}

\bibitem[\protect\citeauthoryear{{Afshordi}, {Loh} \& {Strauss}}{{Afshordi}
  et~al.}{2004}]{Afshordi}
{Afshordi} N.,  {Loh} Y.-S.,    {Strauss} M.~A.,  2004, \prd, 69, 083524

\bibitem[\protect\citeauthoryear{{Bennett} et~al.,}{{Bennett}
  et~al.}{2003}]{Bennett_WMAP1_MNRAS}
{Bennett} C.~L.,  et~al., 2003, \apjs, 148, 1

\bibitem[\protect\citeauthoryear{{Blake}, {Collister}, {Bridle} \&
  {Lahav}}{{Blake} et~al.}{2007}]{Blake_SDSSLRG}
{Blake} C.,  {Collister} A.,  {Bridle} S.,    {Lahav} O.,  2007, \mnras, 374,
  1527

\bibitem[\protect\citeauthoryear{{Boughn} \& {Crittenden}}{{Boughn} \&
  {Crittenden}}{2005}]{Boughn_Crittenden_2005}
{Boughn} S.~P.,  {Crittenden} R.~G.,  2005, New Astronomy Review, 49, 75

\bibitem[\protect\citeauthoryear{{Cabr{\'e}}, {Fosalba}, {Gazta{\~n}aga} \&
  {Manera}}{{Cabr{\'e}} et~al.}{2007}]{Cabre_2007}
{Cabr{\'e}} A.,  {Fosalba} P.,  {Gazta{\~n}aga} E.,    {Manera} M.,  2007,
  \mnras, 381, 1347

\bibitem[\protect\citeauthoryear{{Coles} \& {Jones}}{{Coles} \&
  {Jones}}{1991}]{Coles_Jones}
{Coles} P.,  {Jones} B.,  1991, \mnras, 248, 1

\bibitem[\protect\citeauthoryear{{Cooray} \& {Sheth}}{{Cooray} \&
  {Sheth}}{2002}]{Cooray_and_Sheth2002}
{Cooray} A.,  {Sheth} R.,  2002, Physics Reports, 372, 1

\bibitem[\protect\citeauthoryear{{Crittenden} \& {Turok}}{{Crittenden} \&
  {Turok}}{1996}]{Crittenden_Turok_1996}
{Crittenden} R.~G.,  {Turok} N.,  1996, Physical Review Letters, 76, 575

\bibitem[\protect\citeauthoryear{{Efstathiou} et~al.,}{{Efstathiou}
  et~al.}{2002}]{GPE_LAM02}
{Efstathiou} G.,  et~al., 2002, \mnras, 330, L29

\bibitem[\protect\citeauthoryear{{Efstathiou}, {Sutherland} \&
  {Maddox}}{{Efstathiou} et~al.}{1990}]{GPE_LAM90}
{Efstathiou} G.,  {Sutherland} W.~J.,    {Maddox} S.~J.,  1990, \nat, 348, 705

\bibitem[\protect\citeauthoryear{{Eisenstein} et~al.,}{{Eisenstein}
  et~al.}{2005}]{Eisenstein_BAO_MNRAS}
{Eisenstein} D.~J.,  et~al., 2005, \apj, 633, 560

\bibitem[\protect\citeauthoryear{{Fosalba}, {Gazta{\~n}aga} \&
  {Castander}}{{Fosalba} et~al.}{2003}]{Fosalba}
{Fosalba} P.,  {Gazta{\~n}aga} E.,    {Castander} F.~J.,  2003, \apjl, 597, L89

\bibitem[\protect\citeauthoryear{{Giannantonio}, {Crittenden}, {Nichol},
  {Scranton}, {Richards}, {Myers}, {Brunner}, {Gray}, {Connolly} \&
  {Schneider}}{{Giannantonio} et~al.}{2006}]{Giannantonio_2006}
{Giannantonio} T.,  {Crittenden} R.~G.,  {Nichol} R.~C.,  {Scranton} R.,
  {Richards} G.~T.,  {Myers} A.~D.,  {Brunner} R.~J.,  {Gray} A.~G.,
  {Connolly} A.~J.,    {Schneider} D.~P.,  2006, \prd, 74, 063520

\bibitem[\protect\citeauthoryear{{Giannantonio}, {Scranton}, {Crittenden},
  {Nichol}, {Boughn}, {Myers} \& {Richards}}{{Giannantonio}
  et~al.}{2008}]{Giannantonio_2008}
{Giannantonio} T.,  {Scranton} R.,  {Crittenden} R.~G.,  {Nichol} R.~C.,
  {Boughn} S.~P.,  {Myers} A.~D.,    {Richards} G.~T.,  2008, \prd, 77, 123520

\bibitem[\protect\citeauthoryear{{G{\'o}rski}, {Hivon}, {Banday}, {Wandelt},
  {Hansen}, {Reinecke} \& {Bartelmann}}{{G{\'o}rski}
  et~al.}{2005}]{Gorski_HEALPix}
{G{\'o}rski} K.~M.,  {Hivon} E.,  {Banday} A.~J.,  {Wandelt} B.~D.,  {Hansen}
  F.~K.,  {Reinecke} M.,    {Bartelmann} M.,  2005, \apj, 622, 759

\bibitem[\protect\citeauthoryear{{Granett}, {Neyrinck} \& {Szapudi}}{{Granett}
  et~al.}{2009}]{Granett_ISW09}
{Granett} B.~R.,  {Neyrinck} M.~C.,    {Szapudi} I.,  2009, \apj, 701, 414

\bibitem[\protect\citeauthoryear{{Hambly}, {Davenhall}, {Irwin} \&
  {MacGillivray}}{{Hambly} et~al.}{2001}]{Hambly_superCOSMOS}
{Hambly} N.~C.,  {Davenhall} A.~C.,  {Irwin} M.~J.,    {MacGillivray} H.~T.,
  2001, \mnras, 326, 1315

\bibitem[\protect\citeauthoryear{{Hinshaw} et~al.,}{{Hinshaw}
  et~al.}{2007}]{Hinshaw_WMAP3}
{Hinshaw} G.,  et~al., 2007, \apjs, 170, 288

\bibitem[\protect\citeauthoryear{{Hu} \& {Dodelson}}{{Hu} \&
  {Dodelson}}{2002}]{Hu_Dodelson_CMB}
{Hu} W.,  {Dodelson} S.,  2002, \araa, 40, 171

\bibitem[\protect\citeauthoryear{{Jarrett}}{{Jarrett}}{2004}]{Jarrett_2MASS}
{Jarrett} T.,  2004, Publications of the Astronomical Society of Australia, 21,
  396

\bibitem[\protect\citeauthoryear{{Jeffreys}}{{Jeffreys}}{1948}]{Jeffreys_1948}
{Jeffreys} H.,  1948, Theory of probability.
Oxford: Clarendon Press

\bibitem[\protect\citeauthoryear{{Kaiser}}{{Kaiser}}{1992}]{Kaiser_1992}
{Kaiser} N.,  1992, \apj, 388, 272

\bibitem[\protect\citeauthoryear{{Lewis}, {Challinor} \& {Lasenby}}{{Lewis}
  et~al.}{2000}]{Lewis_CAMB}
{Lewis} A.,  {Challinor} A.,    {Lasenby} A.,  2000, \apj, 538, 473

\bibitem[\protect\citeauthoryear{{Loveday}, {Maddox}, {Efstathiou} \&
  {Peterson}}{{Loveday} et~al.}{1995}]{Loveday_1995}
{Loveday} J.,  {Maddox} S.~J.,  {Efstathiou} G.,    {Peterson} B.~A.,  1995,
  \apj, 442, 457

\bibitem[\protect\citeauthoryear{{Magliocchetti}, {Bagla}, {Maddox} \&
  {Lahav}}{{Magliocchetti} et~al.}{2000}]{Magliocchetti}
{Magliocchetti} M.,  {Bagla} J.~S.,  {Maddox} S.~J.,    {Lahav} O.,  2000,
  \mnras, 314, 546

\bibitem[\protect\citeauthoryear{{Martinez-Gonzalez}, {Sanz} \&
  {Silk}}{{Martinez-Gonzalez} et~al.}{1990}]{Martinez-Gonzalez_Sanz_Silk}
{Martinez-Gonzalez} E.,  {Sanz} J.~L.,    {Silk} J.,  1990, \apjl, 355, L5

\bibitem[\protect\citeauthoryear{{Nolta}, {Wright}, {Page}, {Bennett},
  {Halpern}, {Hinshaw}, {Jarosik}, {Kogut}, {Limon}, {Meyer}, {Spergel},
  {Tucker} \& {Wollack}}{{Nolta} et~al.}{2004}]{Nolta}
{Nolta} M.~R.,  {Wright} E.~L.,  {Page} L.,  {Bennett} C.~L.,  {Halpern} M.,
  {Hinshaw} G.,  {Jarosik} N.,  {Kogut} A.,  {Limon} M.,  {Meyer} S.~S.,
  {Spergel} D.~N.,  {Tucker} G.~S.,    {Wollack} E.,  2004, \apj, 608, 10

\bibitem[\protect\citeauthoryear{{Park}, {Vogeley}, {Geller} \&
  {Huchra}}{{Park} et~al.}{1994}]{Park_1994}
{Park} C.,  {Vogeley} M.~S.,  {Geller} M.~J.,    {Huchra} J.~P.,  1994, \apj,
  431, 569

\bibitem[\protect\citeauthoryear{{Peacock} et~al.,}{{Peacock}
  et~al.}{2010}]{Peacock_photoz}
{Peacock} J.~A.,  et~al., 2010, in prep.

\bibitem[\protect\citeauthoryear{{Percival}, {Cole}, {Eisenstein}, {Nichol},
  {Peacock}, {Pope} \& {Szalay}}{{Percival} et~al.}{2007}]{WJP_BAO07}
{Percival} W.~J.,  {Cole} S.,  {Eisenstein} D.~J.,  {Nichol} R.~C.,  {Peacock}
  J.~A.,  {Pope} A.~C.,    {Szalay} A.~S.,  2007, \mnras, 381, 1053

\bibitem[\protect\citeauthoryear{{Perlmutter} et~al.,}{{Perlmutter}
  et~al.}{1999}]{Perlmutter_MNRAS}
{Perlmutter} S.,  et~al., 1999, \apj, 517, 565

\bibitem[\protect\citeauthoryear{{Pietrobon}, {Balbi} \&
  {Marinucci}}{{Pietrobon} et~al.}{2006}]{Pietrobon_ISW06}
{Pietrobon} D.,  {Balbi} A.,    {Marinucci} D.,  2006, \prd, 74, 043524

\bibitem[\protect\citeauthoryear{{Rassat}, {Land}, {Lahav} \&
  {Abdalla}}{{Rassat} et~al.}{2007}]{Rassat07}
{Rassat} A.,  {Land} K.,  {Lahav} O.,    {Abdalla} F.~B.,  2007, \mnras, 377,
  1085

\bibitem[\protect\citeauthoryear{{Riess} et~al.,}{{Riess}
  et~al.}{1998}]{Reiss_MNRAS}
{Riess} A.~G.,  et~al., 1998, \aj, 116, 1009

\bibitem[\protect\citeauthoryear{{Sawangwit}, {Shanks}, {Cannon}, {Croom},
  {Ross} \& {Wake}}{{Sawangwit} et~al.}{2010}]{Sawangwit_ISW10}
{Sawangwit} U.,  {Shanks} T.,  {Cannon} R.~D.,  {Croom} S.~M.,  {Ross} N.~P.,
   {Wake} D.~A.,  2010, \mnras, 402, 2228

\bibitem[\protect\citeauthoryear{{Schlegel}, {Finkbeiner} \&
  {Davis}}{{Schlegel} et~al.}{1998}]{Schlegel_dust}
{Schlegel} D.~J.,  {Finkbeiner} D.~P.,    {Davis} M.,  1998, \apj, 500, 525

\bibitem[\protect\citeauthoryear{Scranton et~al.,}{Scranton
  et~al.}{2003}]{Scranton_2003}
Scranton R.,  et~al., 2003, astro-ph/0307335

\bibitem[\protect\citeauthoryear{{Spergel} et~al.,}{{Spergel}
  et~al.}{2007}]{Spergel_WMAP3}
{Spergel} D.~N.,  et~al., 2007, \apjs, 170, 377

\end{thebibliography}




\bsp

\label{lastpage}

\end{document}